\documentclass[twocolumn,superscriptaddress,nofootinbib,preprintnumbers,amsmath,amssymb,floatfix,showpacs]{revtex4-1}

\usepackage{graphicx}
\usepackage{amssymb}
\usepackage{dcolumn}
\usepackage{bm}
\usepackage{epsfig}
\usepackage{setspace}
\usepackage[caption=false]{subfig}
\usepackage{xcolor}
\usepackage{changes}

\begin{document}

\title{Modeling the diffusive dynamics of critical fluctuations near the QCD critical point}

\author{Marlene Nahrgang}
\email{marlene.nahrgang@subatech.in2p3.fr}
\affiliation{SUBATECH UMR 6457 (IMT Atlantique, Universit\'e de Nantes,
IN2P3/CNRS), 4 rue Alfred Kastler, 44307 Nantes, France}

\author{Marcus Bluhm}
\email{marcus.bluhm@subatech.in2p3.fr}
\affiliation{SUBATECH UMR 6457 (IMT Atlantique, Universit\'e de Nantes,
IN2P3/CNRS), 4 rue Alfred Kastler, 44307 Nantes, France}

\begin{abstract}
The experimental search for the QCD critical point by means of relativistic heavy-ion collisions necessitates the development of dynamical models of fluctuations. In this work we study the fluctuations of the net-baryon density near the critical point. Due to net-baryon number conservation the correct dynamics is given by the fluid dynamical diffusion equation, which we extend by a white noise stochastic term to include intrinsic fluctuations. We quantify finite resolution and finite size effects by comparing our numerical results to analytic expectations for the structure factor and the equal-time correlation function. In small systems the net-baryon number conservation turns out to be quantitatively and qualitatively important, as it introduces anticorrelations at larger distances. Including nonlinear coupling terms in the form of a Ginzburg-Landau free energy functional we observe non-Gaussian fluctuations quantified by the excess kurtosis. We study the dynamical properties of the system close to equilibrium, for a sudden quench in temperature and a Hubble-like temperature evolution. In the real-time dynamical systems we find the important dynamical effects of critical slowing down, weakening of the extremal value and retardation of the fluctuation signal. In this work we establish a set of general tests, which should be met by any model propagating fluctuations, including upcoming $3+1$ dimensional fluctuating fluid dynamics.
\end{abstract}


\maketitle
 
\section{Introduction}

Conventional fluid dynamics propagates averages of conserved thermodynamic quantities, like the energy density or charge densities, requiring approximate local thermal equilibrium~\cite{Schaefer:2014awa}. Small deviations from equilibrium are described by dissipative corrections, which are quantified by the shear and bulk viscosities and the charge conductivities or diffusion coefficients. In linear response theory these transport coefficients are related to correlators of the fluctuations of thermodynamic quantities in the fluid dynamical limit~\cite{Kovtun:2012rj,Jeon:2015dfa}. By the fluctuation-dissipation theorem it is consistent to not only include the dissipative corrections into the nonlinear fluid dynamical equations of motion but also the propagation of the corresponding intrinsic fluid dynamical fluctuations. These intrinsic fluctuations lead, for example, to non-analytic contributions to the time-dependence of correlations~\cite{Kovtun:2012rj,Kovtun:2003vj,Akamatsu:2016llw,Martinez:2018wia,An:2019osr,An:2019csj}. But most importantly, they become especially interesting when we study the fluid dynamical behavior of a system close to a second-order phase transition~\cite{Hohenberg:1977ym,Son:2004iv,Fujii:2004jt}.

Developing models and simulations for the real-time dynamics  of fluctuations at a phase transition has become increasingly important in the field of relativistic heavy-ion collisions. These are performed experimentally at the Large Hadron Collider (LHC) at CERN, the Relativistic Heavy-Ion Collider (RHIC) at BNL, the Super Proton Synchrotron (SPS) at CERN or the Heavy Ion Synchrotron SIS18 at GSI. In the heavy-ion collisions strongly interacting matter at extreme temperatures $T$ and densities is created~\cite{Jacak:2012dx,Braun-Munzinger:2015hba,Busza:2018rrf}. The successful description of collective effects by conventional fluid dynamical simulations~\cite{Teaney:2009qa,Schenke:2010nt,Heinz:2013th,DelZanna:2013eua,Karpenko:2013wva,deSouza:2015ena,Romatschke:2017ejr} and the modification of high-energetic probes measured in heavy-ion collisions compared to proton-proton collisions~\cite{Connors:2017ptx} are convincing indications for the formation of a new state of matter, the quark-gluon plasma (QGP). At the highest beam energies $\sqrt{s_{\rm NN}}$ at the LHC the QGP is almost baryon free, i.e.~the baryo-chemical potential $\mu_B\simeq 0$, and the transition to hadronic matter is a crossover as demonstrated by lattice QCD calculations~\cite{Aoki:2006we}. As the beam energy is lowered, the phase diagram of QCD can be probed at finite net-baryon density~\cite{Aggarwal:2010cw,Friman:2011zz,Luo:2015doi,Bzdak:2019pkr,Luo:2020pef}. An especially interesting region in the phase diagram is associated with the conjectured critical point beyond which the transition to hadronic matter turns into a first-order phase transition~\cite{Rajagopal:1992qz,Berges:1998rc,Halasz:1998qr,Fukushima:2010bq,Fukushima:2013rx}. Near the critical point fluctuations in conserved charges are expected to grow large and to imprint on the experimentally observed particle multiplicities in form of large event-by-event fluctuations~\cite{Stephanov:1998dy,Stephanov:1999zu,Hatta:2003wn,Asakawa:2015ybt,Luo:2017faz}. Indeed, first measurements during the beam energy scan phase I at RHIC and by the HADES experiment at GSI have shown interesting features in the kurtosis, a fluctuation measure associated with the fourth-order cumulant, of the net-proton distribution~\cite{Adamczyk:2013dal,Adam:2020unf,Adamczewski-Musch:2020slf}. In thermodynamic, i.e.~static and infinite, systems these higher-order cumulants are known to be in particular sensitive to the growth of the correlation length of the associated critical fluctuations~\cite{Stephanov:2008qz,Asakawa:2009aj,Stephanov:2011pb}. 

Up to this day it is unknown quantitatively how critical fluctuations develop in real-time dynamics. Qualitatively, dynamical fluctuations of the chiral condensate or the net-baryon density, as two possible order parameters, have been studied in various works~\cite{Berdnikov:1999ph,Nahrgang:2011mg,Nahrgang:2011mv,Nahrgang:2011vn,Herold:2013bi,Nahrgang:2013jx,Herold:2014zoa,Mukherjee:2015swa,Herold:2016uvv,Nahrgang:2016eou,Mukherjee:2016kyu,Herold:2017day,Stephanov:2017ghc,Herold:2018ptm,Rajagopal:2019xwg,Du:2020bxp,Kitazawa:2013bta,Sakaida:2014pya,Sakaida:2017rtj,Nahrgang:2017hkh,Nahrgang:2018afz,Bluhm:2018qkf,Akamatsu:2018vjr,Bluhm:2019yfb}. The lack of a more quantitative description is mainly due to the challenges that have to be met when including fluctuations in to the standard models of heavy-ion collisions, see~\cite{Bluhm:2020mpc} for a recent review. For the fluid dynamical description it is rather straightforward to include criticality on the level of the equation of state~\cite{Nonaka:2004pg,Bluhm:2006av,Parotto:2018pwx}, but the formulation of algorithms to treat intrinsic fluctuations in this framework remains a challenge~\cite{Young:2013fka,Murase:2016rhl,Nahrgang:2017oqp,Bluhm:2018plm,Singh:2018dpk,Hirano:2018diu,Sakai:2020pjw,2007PhRvE..76a6708B,2009arXiv0906.2425D,delaTorre:2014mys}. For the microscopic transport models, where fluctuations are inherently present, the inclusion of a critical point remains complicated. 

In this work we study the dynamics of fluctuations in a simpler fluid dynamical model, the diffusion equation in one spatial dimension. Our main intent is to report the development of an algorithm, which treats fluctuations for the crucial long-wavelength modes reliably, and to present corresponding benchmark tests that should be met by all future approaches that deal with fluid dynamical fluctuations. 
We focus on the net-baryon density, which in the long-time limit becomes the critical mode associated with the critical point in QCD. We include the critical physics in the vicinity of the QCD critical point by a Ginzburg-Landau free energy functional, motivated by the $3$D Ising universality class. We then test the presented algorithm for the linear Gaussian limits in equilibrium. Here, in particular the static structure factor and the equal-time  correlation function are useful quantities for probing the dynamics of the fluctuations. We then evaluate the dynamical properties of the system, by looking at the dynamic structure factor in equilibrium first. Here, we recover the expected dynamical universality class of model B~\cite{Hohenberg:1977ym}. Next, we investigate the scenario of a sudden temperature-quench and finally a Hubble-like evolution of the temperature. We observe effects of critical slowing down, a weakening and a retardation of the maximal signal.

\section{Diffusive dynamics near the QCD critical point}
\label{sec:model}

The equations of relativistic fluid dynamics describe the conservation of energy and momentum and of net-charges via 
\begin{align}
 \partial_\mu T^{\mu\nu} &= 0 \,,\\
 \partial_\mu N^\mu_{i} &= 0\, .
\end{align}
For our purpose we focus on the non-relativistic evolution of the net-baryon number current 
$N^\mu_{B}=n_B u^\mu + j_B^\mu$, where the Navier-Stokes expression for the viscous current is given by
\begin{equation}
 j_B^\mu = -\Gamma T\Delta^{\mu\nu}\partial_\nu\left(\frac{\mu_B}{T}\right)
\end{equation}
with $\Delta^{\mu\nu}=u^\mu u^\nu-g^{\mu\nu}$, fluid velocity $u^\mu$ and mobility coefficient $\Gamma$. We consider a system that is decoupled from the fluid velocity field which we assume to be space-time independent. In this case we recover the diffusion equation
\begin{equation}
 \partial_t n_B = \Gamma T\nabla^2\left(\frac{\mu_B}{T}\right)
\end{equation}
for the net-baryon density $n_B$. 
The diffusive dynamics happens such as to minimize the free energy in the system. With the thermodynamic relation $\mu_{B}=\delta {\cal F}/\delta n_B$ one obtains the diffusion equation generated by the variation of the free energy functional ${\cal F}$ for a system of spatially homogeneous temperature
\begin{equation}
 \partial_t n_B = \Gamma \nabla^2\bigg(\frac{\delta {\cal F}[n_B]}{\delta n_B}\bigg)\,.
\end{equation}

Since we are interested in the dynamics of intrinsic fluctuations near the critical point we include a stochastic term to arrive at the stochastic diffusion equation
\begin{equation}
 \partial_t n_B = \Gamma \nabla^2\bigg(\frac{\delta {\cal F}[n_B]}{\delta n_B}\bigg) + \vec{\nabla}\cdot\vec{J} \,,
 \label{eq:diffeq1}
\end{equation}
where $\vec{J}$ is a stochastic current given by
\begin{equation}
 \vec{J} = \sqrt{2 T \Gamma} \vec{\zeta}
\end{equation}
and $\vec{\zeta}$ is a Gaussian spatio-temporal white noise field with zero mean and unit variance. Fulfilling the fluctuation-dissipation theorem the covariance of the stochastic term guarantees that the long-time equilibrium distribution is given by
\begin{equation}
P_{\rm eq}[n_B] = \frac{1}{{\cal Z}}\exp\bigg(\frac{-{\cal F}[n_B]}{T}\bigg)\, , 
\end{equation}
normalized by the partition function ${\cal Z}$. 

We choose the free energy functional near the QCD critical point to be of the following polynomial form in $\Delta n_B = n_B-n_c$ with critical density $n_c$: 
\begin{multline}
 {\cal F}[n_B] = T\int{\rm d}^3 x \left(\frac{m^2}{2n_c^2}(\Delta n_B)^2 + \frac{K}{2n_c^2} 
 (\nabla n_B)^2 + \right. \\
 \left. \frac{\lambda_3}{3n_c^3}(\Delta n_B)^3 + \frac{\lambda_4}{4n_c^4}(\Delta n_B)^4 + \frac{\lambda_6}{6n_c^6}(\Delta n_B)^6 \right) \,.
\label{eq:GLpotential}
\end{multline}
We note that the chosen Ginzburg-Landau form for the critical part of the free energy ${\cal F}$ may be augmented by regular contributions. The coupling coefficients can be calculated through the mapping of the $3$-dimensional Ising spin model onto a universal effective potential~\cite{Tsypin:1994nh,Tsypin:1997zz}. This determines the dependence of these couplings on the thermodynamic correlation length $\xi$ within the given universality class as 
\begin{align}
 \label{eq:couplingsA}
 m^2 &= \frac{1}{\xi_0\xi^2}\, ,\\
 K &= \tilde K/\xi_0\, ,\\
 \lambda_3 &= n_c\, \tilde\lambda_3\, (\xi/\xi_0)^{-3/2}\, ,\\
 \lambda_4 &= n_c\, \tilde\lambda_4\,(\xi/\xi_0)^{-1}\,, \\
 \lambda_6 &= n_c\, \tilde\lambda_6\, .
 \label{eq:couplings}
\end{align}
In principle, the dimensionless couplings $\tilde\lambda_3$, $\tilde\lambda_4$ and $\tilde\lambda_6$ have universal values as well, but the uncertainty in translating the spin variables to the QCD phase diagram leads to rather unknown values for these couplings. We will use $\tilde\lambda_3=1$, $\tilde\lambda_4=10$ and $\tilde\lambda_6=3$ in this work. This implies that the temperature dependence of the couplings is determined entirely by the behavior of $\xi$ established through a matching to the susceptibility of the Ising model scaling equation of state~\cite{Guida:1996ep}. In the work~\cite{Tsypin:1994nh,Tsypin:1997zz} it turned out to be important to include the $\lambda_6$ coupling in order to describe the probability distribution of the fluctuations in the spin model. We therefore include this term in our study as well, although in a perturbative expansion in $\xi^3/V$ with volume $V$ this term is suppressed in the scaling regime~\cite{Stephanov:2008qz,Nouhou:2019nhe}. 

\begin{figure}
 \centering
 \includegraphics[width=0.48\textwidth]{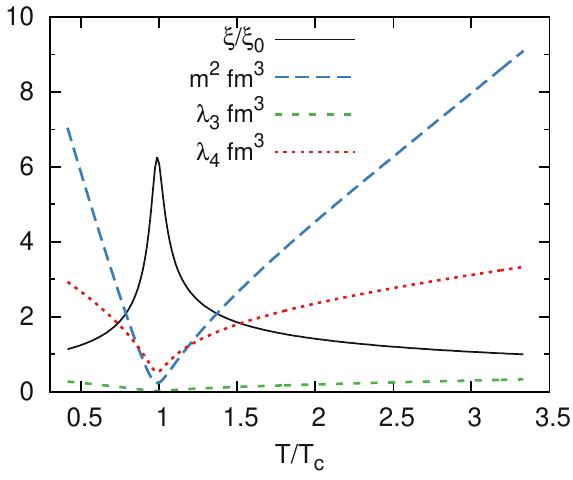}
 \caption{Scaled temperature dependence of the parameters in the Ginzburg-Landau free energy functional ${\cal F}$ in Eq.~(\ref{eq:GLpotential}). We choose $T_c=0.15$~GeV and $\xi=\xi_0=0.479$~fm at $T=T_0=0.5$~GeV. Furthermore, the coupling $\lambda_6=1/{\rm fm}^3$ (not shown) is set constant as a function of $T$. The temperature dependence of $\xi/\xi_0$, which serves as input for the parameters in this work, follows from a matching to the susceptibility of the Ising model scaling equation of state for constant $\mu_B$ on the crossover side of the QCD phase diagram, see~\cite{Bluhm:2016trm,Bluhm:2016byc} for some details.}
 \label{fig:parameter}
\end{figure}

As can be seen in Fig. \ref{fig:parameter}, the thermodynamic correlation length peaks around $T_c$ which we choose as $T_c=0.15$~GeV, while the couplings $\lambda_3$ and $\lambda_4$ have a minimum at $T_c$. There is a region around $T_c$ where the nonlinear couplings $\lambda_4$ and $\lambda_6$ are larger than the Gaussian mass parameter $m$. We expect nonlinear effects to be largest here. The critical net-baryon density $n_c$ depends on the location of the critical point and the equation of state. The net-baryon density at chemical freeze-out as a function of $\sqrt{s_{\rm NN}}$ was obtained from statistical model fits using the Hadron Resonance Gas model in~\cite{Randrup:2009ch}. Here, maximal values of $n_B=0.12/$fm$^3$ are reached at $\sqrt{s_{\rm NN}}\sim 4$~GeV. During the evolution the system can reach much higher local values of $n_B=5\rho_0$ with $\rho_0=0.16/$fm$^3$~\cite{Bravina:2008ra}. In this work, we choose a value of $n_c=1/(3{\rm fm}^3)$. 

The above described setup is in general designed for studying the diffusion dynamics of critical fluctuations in three spatial dimensions. The numerical framework presented here focusses on the dynamics restricted to one spatial direction. For this purpose, we scale out the transverse area $A$ and consider the dynamics only in the longitudinal direction which resembles the situation met in a highly anisotropic heavy-ion collision. With this, the stochastic diffusion equation Eq.~(\ref{eq:diffeq1}) becomes
\begin{multline}
 \partial_t n_B(x,t) = \frac{D}{n_c} \left(m^2 \nabla_x^2 n_B - K\nabla_x^4 n_B\right)\\ + D\nabla_x^2\left(\frac{\lambda_3}{n_c^2}\, (\Delta n_B)^2 + \frac{\lambda_4}{n_c^3}\, (\Delta n_B)^3 + \frac{\lambda_6}{n_c^5}\, (\Delta n_B)^5\right) \\+ \sqrt{2Dn_c/A}\,\nabla_x\zeta_x(x,t) \,,
\label{eq:stochdiff}
\end{multline}
where we have expressed the mobility coefficient $\Gamma = D n_c/T$ via the diffusion coefficient $D$ and the covariance reads $\langle\zeta_x(x,t),\zeta_x(x',t')\rangle=\delta(x-x')\delta(t-t')$.

\section{Equilibrium fluctuations}
\label{sec:diffequi}

In this section we investigate the long-time limit for the stochastic diffusion of the net-baryon density at various fixed thermal conditions. For this purpose, we consider a system in a quasi one-dimensional box of length $L$ with periodic boundary conditions. Initially, the net-baryon density is constant and set to $n_B(x) = n_{c}$. Both, the discretization with $\Delta x = L/N_x$ ($N_x$ is the number of sites) and the finite size of the box will introduce effects which make the results differ from the continuum limit ($\Delta x\to 0$) and the thermodynamic limit ($L\to\infty$). While the limited resolution is a technical issue, the finite size reflects the situation of the fireball created in a heavy-ion collision. After initialization we let the system equilibrate during a long time, which is proportional to $L^2/D$, before evaluating the physical observables such as the variance and kurtosis or the equal-time corelation function and structure factor of the system. These are related to the equilibrium distribution which is an invariant measure and independent of $D$. The latter is exemplarily set to $D=1$~fm. 

We note that the determination of equilibrium results, i.e.~the long-time behavior, numerically requires a significant amount of statistics. For dissipation in form of diffusion any memory on initial conditions is eventually lost and the fluctuation-dissipation balance guarantees ergodicity of the system. This implies that ensemble averages can be either obtained by averaging over multiple samples or equally by averaging over time after performing a sufficient amount of equilibration steps proportional to $L^2/(D\Delta t)$. In this work, the high-statistics equilibrium results have been obtained by combining both methods. 

We solve the stochastic diffusion equation Eq.~(\ref{eq:stochdiff}) numerically within a semi-implicit scheme, where the nonlinear terms in $\Delta n_B$ are treated explicitly. Charge conservation is respected with very high precision by imposing periodic boundary conditions. More details can be found in Appendix~\ref{app:A}.

\subsection{Static structure factor and equal-time correlation function in Gaussian models}
\label{sec:staticSF}

The stochastic diffusion equation Eq.~(\ref{eq:stochdiff}) contains different physics cases. For the Gaussian models the nonlinear couplings $\lambda_i$ are equal to zero. In this case, exact analytic continuum expressions for prominent physical observables are calculable. One of these represents the dynamic structure factor $S(k,\omega)$ for wavevector $\vec{k}$ and frequency $\omega$. It follows directly from the space-time Fourier transform of the stochastic diffusion equation as 
\begin{align}
\nonumber
 S(k,\omega)\equiv & \,V\langle \Delta \hat{n}_B(k,\omega)\,\Delta \hat{n}_B^*(k,\omega)\rangle \\
 = & \,\frac{2Dn_ck^2}{\omega^2+\left[Dk^2(m^2+Kk^2)/n_c\right]^2}
\label{equ:dynamicSdefinition}
\end{align}
and entails the dynamical space-time spectrum of the fluctuating net-baryon density. We note that for the spatio-temporal white noise field the dynamic structure factor is $S_{\zeta_x}(k,\omega)=L\,\langle\hat{\zeta_x}\,\hat{\zeta_x}^*\rangle=1$. From $S(k,\omega)$ the spatial spectrum at equal time, i.e.~the static structure factor $S(k)$, follows from integration over all $\omega$ as
\begin{equation}
\label{eq:definitionSstatic}
 S(k)=\frac{1}{2\pi}\int_{-\infty}^{\infty} S(k,\omega)\,d\omega\,.
\end{equation}

The simplest version of a Gaussian model is obtained when $\tilde K=\tilde \lambda_3=\tilde \lambda_4 =\tilde \lambda_6 = 0$ in Eq.~(\ref{eq:GLpotential}). In this case we are left with the Gaussian mass term which gives rise to the standard diffusion equation. This model serves as a reference and was discussed in detail in~\cite{Nahrgang:2017hkh}, where the correct numerical implementation of Eq.~\eqref{eq:stochdiff} for this case was verified. From Eq.~\eqref{equ:dynamicSdefinition} the static structure factor for $\tilde{K}=0$ follows via Eq.~\eqref{eq:definitionSstatic} as
\begin{equation}
 S(k) = \frac{n_c^2}{m^2}\,,
\label{eq:SkGausscont}
\end{equation}
which is independent of the wavevector $\vec{k}$. Contrary to simple Euler schemes, the semi-implicit scheme applied in our framework achieves highest accuracy for all wavenumbers independent of the time step $\Delta t$. As we show in Appendix~\ref{app:A+}, the corresponding structure factor in discretized space-time $S_k$ coincides with Eq.~(\ref{eq:SkGausscont}) and is therefore independent of the lattice spacing $\Delta x$. In~\cite{Nahrgang:2017hkh} we verified that this is reproduced in our framework.

The version with a term of non-zero $\tilde{K}$, which describes a kinetic energy in a Klein-Gordon type action or a surface tension in diffusion equations, can still be solved analytically in the continuum. In this case, which we will call Gauss$+$surface model, the static structure factor is given by 
\begin{equation}
 S(k) = \frac{n_c^2}{m^2} \frac{1}{1 + Kk^2/m^2}\,.
 \label{eq:SkGaKcont}
\end{equation}
Due to the finite surface tension the amplitude of the fluctuations becomes suppressed with increasing $k$. 

\begin{figure}
 \centering
 \includegraphics[width=0.48\textwidth]{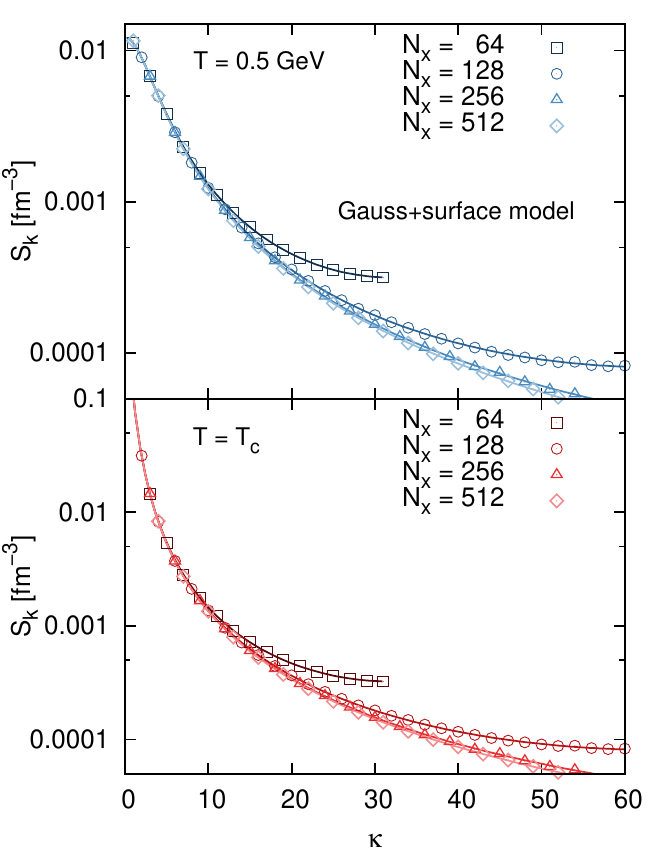}
 \caption{The static structure factor (symbols) as a function of wavenumber $\kappa$ in the Gauss+surface model ($\lambda_3=\lambda_4=\lambda_6=0$) for different $N_x= 64, 128, 256, 512$ and fixed $L=10$~fm. For both temperatures, $T=0.5$~GeV and $T=T_c$, the theoretical expectations (solid curves) for the static structure factor in discretized space-time $S_k$, Eq.~(\ref{equ:StrucFacDiscretGA+K}) with $k=2\pi\kappa/L$, are perfectly reproduced. Because of the Hermitian symmetry between $\kappa$ and $N_x-\kappa$, $S_k$ is symmetric about $\kappa=N_{\rm x}/2$. With increasing resolution $S_k$ approaches the continuum result $S(k)$ in Eq.~\eqref{eq:SkGaKcont}.}
 \label{fig:surfaceSk}
\end{figure}

The numerical results presented in this work have been obtained for $\tilde{K}=1$ in each of the calculations. For our numerical framework, the static structure factor for the Gauss$+$surface model in discretized space-time reads (see Appendix~\ref{app:A+})
\begin{equation}
 \label{equ:StrucFacDiscretGA+K}
 S_k = \frac{n_c^2}{m^2}\frac{1}{1+\frac{2K}{m^2\Delta x^2}\left[1-\cos(k\Delta x)\right]}\, .
\end{equation}
With increasing resolution, $\Delta x \to 0$, this result converges to Eq.~(\ref{eq:SkGaKcont}). In Fig.~\ref{fig:surfaceSk} we show the numerical results for the static structure factor $S_k$ as a function of wavenumber $\kappa$ for fixed box length $L=10$~fm and different resolutions at two different temperatures. As the considered box is finite in size and the resolution limited by $N_x$ only a finite number of modes with discrete $\kappa=kL/(2\pi)$ are realized. Our numerical implementation reproduces the analytic expectations for $S_k$ from Eq.~\eqref{equ:StrucFacDiscretGA+K}, thus, resolution effects are well understood. 
We note that for a resolution of $\Delta x=(10/128)$~fm the static structure factor starts to deviate visibly from the continuum result only for $\kappa\gtrsim 25$ while for the modes $\kappa\lesssim 10$, which are important for the critical physics, the continuum limit is reached. 
Close to $T_c$ the amplitude of fluctuations for modes with small $\kappa$ is increased compared to temperatures further away while $S_k$ is rather independent of $T$ for larger wavenumbers. 

Another prominent observable is the equal-time correlation function of density fluctuations in coordinate space. In the continuum limit it is defined as the Fourier transform of the static structure factor $S(k)$ in Eq.~\eqref{eq:definitionSstatic} via 
\begin{equation}
 \label{equ:FourierTrfContinuum}
 \langle\Delta n_B(r)\Delta n_B(0)\rangle = \int\frac{{\rm d}^dk}{(2\pi)^d} \,e^{i \,\vec{k}\cdot\vec{r}} 
 S(k) \,. 
\end{equation}
For the quasi $d=1$ dimensional system studied in our work the equal-time correlation function of density fluctuations in the longitudinal direction is given for the Gauss$+$surface model by
\begin{equation}
 \label{eq:CorrFunction1D}
 \langle\Delta n_B(r)\Delta n_B(0)\rangle = 
 \frac{{n_c^2}}{2 Am\sqrt{K}} \exp\left(-r\frac{m}{\sqrt{K}}\right) \,. 
\end{equation}
For $\tilde K = 1$ we recover the standard relation between the Gaussian mass parameter and the correlation length given by Klein-Gordon theory. The truly realized correlation length in the system depends, however, in general on the surface tension. The integral of Eq.~(\ref{eq:CorrFunction1D}) over distances much larger than the correlation length yields the full weight of the fluctuation, $n_c^2/(Am^2)$. This is the same as for the pure Gaussian model with vanishing $\tilde K$, where Eq.~\eqref{eq:CorrFunction1D} reduces to $\langle\Delta n_B(r)\Delta n_B(0)\rangle = (n_c^2/(Am^2))\,\delta(r)$ 
and the expected uncorrelated Gaussian limit is recovered. 
\begin{figure}
\centering
 \includegraphics[width=0.48\textwidth]{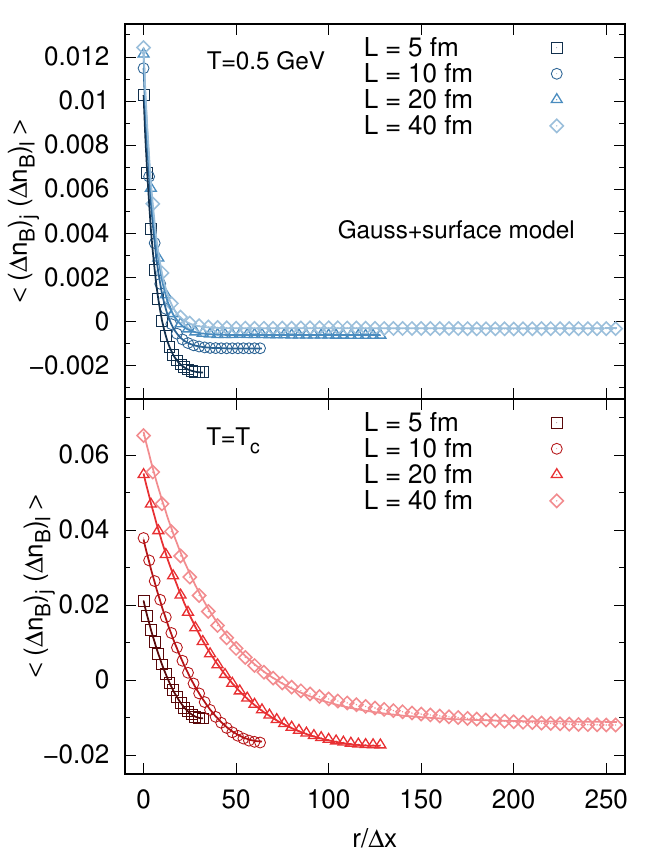}
 \caption{The equal-time correlation function  of density fluctuations (symbols) in the Gauss+surface model ($\lambda_3=\lambda_4=\lambda_6=0$) for different $L/$fm $=5, 10, 20, 40$, fixed $\Delta x=(20/256)$~fm, i.e.~different $N_x$, and a representative $A=1$~fm$^2$. For both temperatures, $T=0.5$~GeV and $T=T_c$, the numerical results are found to perfectly agree with the theoretical expectations (solid curves) when including the finite size corrections for exact net-baryon number conservation, cf.~Appendix~\ref{app:B}. The correlation function is symmetric in $r=|j-l|\Delta x$ about $r=L/2$.}
 \label{fig:surfaceCorrFunc3}
\end{figure}

In~\cite{Nahrgang:2017hkh}, the behavior of $\langle\Delta n_B(r)\Delta n_B(0)\rangle$ for the pure Gaussian model was studied numerically. For this model the correlation function in discretized space-time is given by
$\langle(\Delta n_B)_j(\Delta n_B)_l\rangle = n_c^2/(Am^2 \Delta x)\,\delta_{jl}$, where $j, l$ can go over all cells. Accordingly, fluctuations are uncorrelated over distances larger than the lattice spacing. In our simulations exact net-baryon number conservation is realized over the entire box of finite length $L$. This leads to corrections which can analytically be understood by imposing the condition of charge conservation $\sum_l \langle(\Delta n_B)_j(\Delta n_B)_l\rangle = 0$ for any $j$, see Appendix~\ref{app:B}. Correspondingly, the expectation for the equal-time correlation function changes to 
\begin{equation}
 \label{equ:pureGaussSigma}
 \langle(\Delta n_B)_j(\Delta n_B)_l\rangle = \frac{n_c^2}{Am^2}\left(\frac{\delta_{jl}}{\Delta x} - \frac{1}{L}\right) \,,
\end{equation}
which amounts to a constant negative shift that vanishes with increasing $L$ for fixed resolution. This behavior was found to be perfectly reproduced in the numerics, see~\cite{Nahrgang:2017hkh}. 

For the Gauss$+$surface model similar considerations can be made. Numerical results for the equal-time correlation function $\langle(\Delta n_B)_j(\Delta n_B)_l\rangle$ as a function of scaled distance $r/\Delta x=|j-l|$ are shown in Fig.~\ref{fig:surfaceCorrFunc3} for fixed resolution $\Delta x$ and various $L$ at two different temperatures. We find that the equal-time correlation function is shifted to negative values at large distances $r$ demonstrating significant anticorrelations. With increasing box size $L$ at fixed resolution the negative shift becomes less pronounced. This behavior is a consequence of exact net-baryon number conservation, see Appendix~\ref{app:B}. Taking the latter into account, cf.~Eq.~\eqref{eq:CorrfunCorr}, the corresponding analytic expectations agree well with our numerical results, thus, finite size effects in connection with exact charge conservation are well under control.

For temperatures close to $T_c$, $\langle(\Delta n_B)_j(\Delta n_B)_l\rangle$ becomes broader and correlations form over larger distances as one expects from the continuum expression in Eq.~\eqref{eq:CorrFunction1D}. Nonetheless, this depends strongly on the size of the box and finite-size effects in connection with charge conservation clearly affect the development of the correlations. We note that for the larger systems the equilibration times become very long and increasing computer resources are needed to produce equilibrated systems and build up the expected long-range correlations. In fact, the tiny deviation between theoretical expectations and numerical results at large $r$ seen in Fig.~\ref{fig:surfaceCorrFunc3} at $T=T_c$ for $L=40$~fm is the result of an insufficient equilibration before evaluating the equal-time correlation function.

\subsection{Static structure factor and equal-time correlation function in the Ginzburg-Landau model\label{sec:StaticNonGauss}}

\begin{figure}
 \centering
 \includegraphics[width=0.48\textwidth]{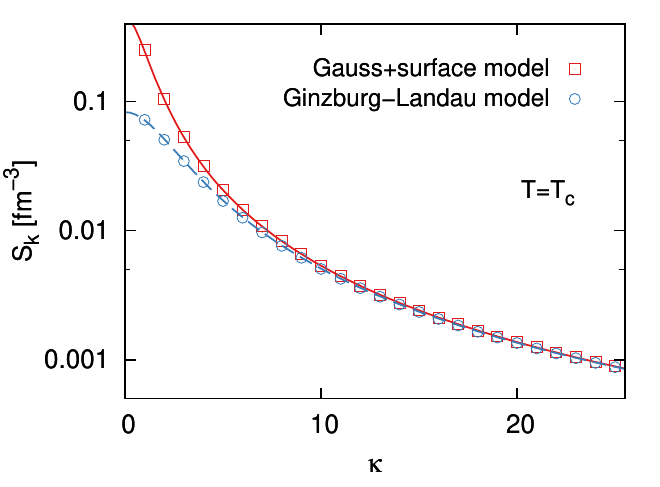}
 \includegraphics[width=0.48\textwidth]{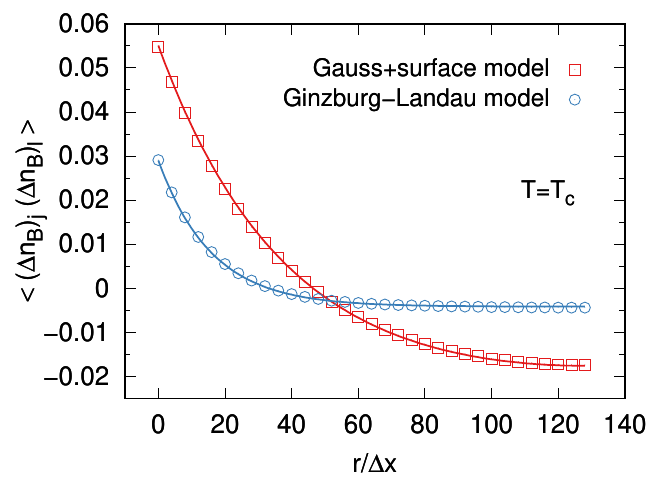}
 \caption{Comparison of the static structure factor (upper panel) and the equal-time correlation function of density fluctuations (lower panel) between Ginzburg-Landau model (circles) and Gauss$+$surface model (squares) for a system of $L=20$~fm, $N_x=256$ and $A=1$~fm$^2$ at $T=T_c$. The theoretical expectations in the Gauss$+$surface model (solid red curves), see Sec.~\ref{sec:staticSF}, agree with the numerics. The numerical results of the Ginzburg-Landau model can formally be described by the same analytic expressions when replacing $m$ by an effective mass $m_{\rm eff}$ that is fitted to describe $S_k$ (dashed blue curves).}
 \label{fig:staticSkCorrGL}
\end{figure}

Let us now study the impact of the nonlinear coupling terms in what we call the Ginzburg-Landau model on the static structure factor and the equal-time correlation function. This is shown in Fig.~\ref{fig:staticSkCorrGL} in comparison with the Gauss$+$surface model results for a system of $L=20$~fm with $N_x=256$ at $T=T_c$. One observes that the influence of the non-zero $\lambda_i$ is the significant reduction of $S_k$ at small wavenumbers $\kappa$ while it is less important for larger $\kappa$. This reduction of the amplitude of fluctuations at long wavelengths is also reflected in the development of spatial correlations. With non-zero $\lambda_i$, the equal-time correlation function is less broad and long-range correlations are suppressed. In addition, correlations at small distances are less pronounced which consequently reduces the quantitative impact of exact charge conservation in the finite-size system. These effects are found to be less important for $T$ further away from $T_c$.

The numerical results of the Gauss$+$surface model can successfully be described by our analytic expectations in discretized space-time, see Sec.~\ref{sec:staticSF}. For the Ginzburg-Landau model, instead, no exact analytic expressions can be derived to compare the numerics with. We note, however, that the numerical results of the Ginzburg-Landau model on the level of $2$-point correlations can formally be described by the analytic expressions of the Gauss$+$surface model but with a modified Gaussian mass parameter while $K$ is kept fixed. This effective mass, $m_{\rm eff}$, is larger than $m$ of the Gauss$+$surface model for any $T$. Near $T_c$ the relative increase of $m_{\rm eff}$ with respect to $m$ is stronger. For the systems studied in this work we find no additional $\Delta x$-dependence in $m_{\rm eff}$ within the statistical uncertainty. 

\subsection{Temperature and system-size dependence of the correlation length\label{sec:NumCorrLength}}

The continuum expectation of the equal-time correlation function in the Gauss$+$surface model for an infinite system is given in Eq.~\eqref{eq:CorrFunction1D}. The numerical results in discretized space resemble this form of an exponential decay. This is also the case when taking non-zero $\lambda_i$ into account. As we have seen in Figs.~\ref{fig:surfaceCorrFunc3} and~\ref{fig:staticSkCorrGL}, net-baryon number conservation in the finite-size system results in a negative offset signalling anticorrelations. Still, an exponential form of the correlation function remains. Therefore, we may fit the numerical results of the Gauss$+$surface and Ginzburg-Landau models with an ansatz that contains the exponential behaviour and the offset (see Appendix~\ref{app:C} for details) in order to determine the correlation length $\tilde\xi$. The latter depends besides $T$ in particular on the system size $L$ and can be different from the thermodynamic correlation length $\xi$.

\begin{figure}
 \centering
 \includegraphics[width=0.48\textwidth]{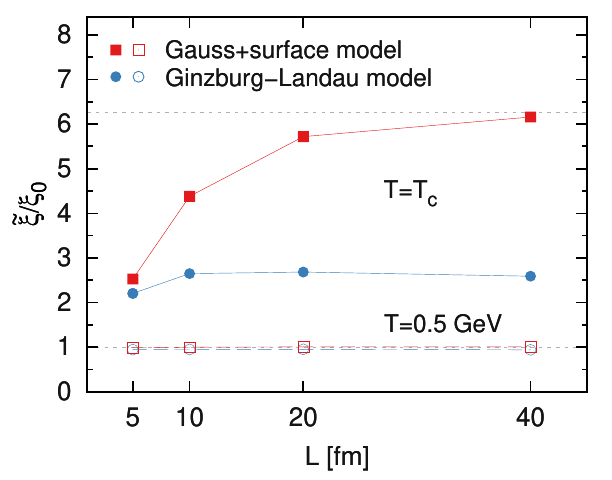}
 \caption{$L$-dependence of the correlation length $\tilde\xi$ in units of $\xi_0$ in the Gauss$+$surface (squares) and Ginzburg-Landau (circles) models at fixed resolution $\Delta x=(20/256)$~fm for two different temperatures, $T=0.5$~GeV (open symbols) and $T=T_c$ (solid symbols). The horizontal, grey dotted lines show for comparison the corresponding scaled thermodynamic correlation length $\xi/\xi_0$ for an infinite system, cf.~the input parameters in Fig.~\ref{fig:parameter}.}
 \label{fig:surfaceCorrFunc}
\end{figure}
In Fig.~\ref{fig:surfaceCorrFunc} we show the system-size dependence of the fitted $\tilde\xi$ in the Gauss$+$surface and Ginzburg-Landau models for two different $T$ at fixed resolution $\Delta x=(20/256)$~fm. The residual $\Delta x$-dependence can be estimated to be on the per cent level for all $T$ and $L$. For the parameters studied in this work, cf.~Fig.~\ref{fig:parameter}, the maximally reached thermodynamic correlation length in an infinite system, $\xi$, is about $3$~fm near $T_c$ and minimally we have $\xi=\xi_0$ at $T=0.5$~GeV. These values are indicated by the grey dotted lines in Fig.~\ref{fig:surfaceCorrFunc}. For $T=0.5$~GeV (open squares and circles) a system size of $L=5$~fm is already sufficient for $\tilde\xi$ to reach approximately the value of $\xi$. This remains unchanged with increasing $L$. However, for all other $T$ with a larger $\xi$ charge conservation turns out to be important, particularly in the smaller systems. In fact, it can lead to a sizeable reduction of $\tilde\xi$ compared to $\xi$ for $L=5$~fm. This effect is pronounced strongest at $T_c$ (solid squares and circles). For $T=T_c$ the fitted correlation length increases strongly toward $\xi$ with increasing $L$ for the Gauss$+$surface model. For $L=40$~fm one finds $\tilde\xi$ to be approximately $\xi$. In contrast, in the Ginzburg-Landau model $\tilde\xi$ remains always small compared to $\xi$ and shows within the statistics a negligible system-size dependence for $L\geq 10$~fm. This reduction is entirely a consequence of the nonlinear interactions.

\begin{figure}
 \centering
 \includegraphics[width=0.48\textwidth]{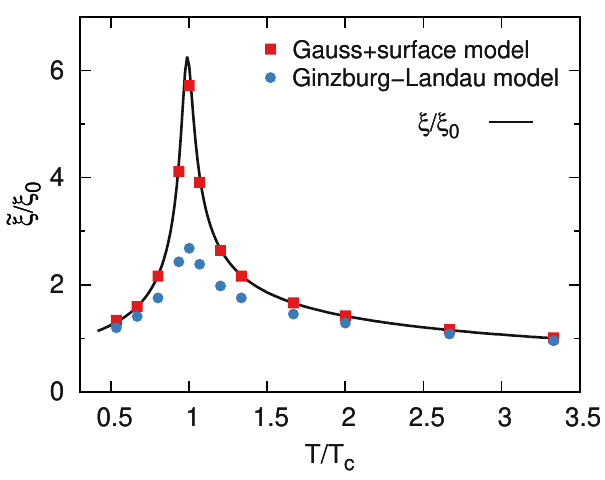}
 \caption{Comparison of the scaled temperature dependence of the correlation length $\tilde\xi$ in units of $\xi_0$ in the Gauss$+$surface (squares) and Ginzburg-Landau (circles) models with the scaled thermodynamic correlation length $\xi/\xi_0$ for an infinite system (solid line) for simulations with $L=20$~fm and $\Delta x=L/256$.}
 \label{fig:XiFit}
\end{figure}
In Fig.~\ref{fig:XiFit} we compare for $L=20$~fm the fitted correlation length as a function of temperature with $\xi$. One observes that $\tilde\xi$ is approximately $\xi$ in the Gauss$+$surface model for all $T$ except very close to $T_c$, where finite-size and charge-conservation effects are strongest, cf.~Fig.~\ref{fig:surfaceCorrFunc}. From this observation we conclude that in order to draw physical conclusions a reasonable compromise between finite-resolution and finite-size effects on the one hand and limited computational resources on the other hand is to study systems of $L=20$~fm and $N_x=256$ in this work. The presence of the nonlinear interactions in the Ginzburg-Landau model impacts the development of long-range correlations significantly. For all $T$ we find a $\tilde\xi$ which is smaller in the Ginzburg-Landau model than in the Gauss$+$surface model. While far away from $T_c$ the effect is tiny, the reduction is visible in the vicinity of $T_c$. This behaviour is in line with the temperature dependence of the parameters, see Fig.~\ref{fig:parameter}, and with the observation that for describing the structure factor and the correlation function in the Ginzburg-Landau model by the analytic expressions of the Gauss$+$surface model one needs $m_{\rm eff}>m$. In fact, we find that $m_{\rm eff}/m$ behaves approximately like the ratio of the fitted correlation lengths in the Gauss$+$surface to the Ginzburg-Landau model. We expect that the fluctuation observables are similarly affected by this.

\subsection{Temperature and system-size dependence of Gaussian and non-Gaussian fluctuations}

We now turn to the study of fluctuation observables in the Gauss$+$surface and Ginzburg-Landau models. We will concentrate on the discussion of local quantities, i.e.~on the fluctuations in the net-baryon density contained within one grid spacing, on an event-by-event basis. The local variance, $\sigma^2$, is equivalent to the equal-time correlation function $\langle(\Delta n_B)^2\rangle$ at $r=0$. From Eq.~\eqref{eq:CorrFunction1D} we see that $\sigma^2\sim\xi$. Since the Gaussian mass parameter $m\sim 1/\xi$ drops rapidly around $T_c$ with a minimum at $T_c$, cf.~Fig.~\ref{fig:parameter}, we expect that the local variance is largest at $T_c$ in both the Gauss$+$surface and the Ginzburg-Landau model. 
The local excess kurtosis, $\kappa$, is defined as 
\begin{equation}
 \kappa=\frac{\mu_4}{\sigma^4}-3 \,,
 \label{eq:kappa}
\end{equation}
where $\mu_4 = \langle (\Delta n_B)^4\rangle$ at $r=0$ is the fourth central moment of local fluctuations. The excess kurtosis must vanish for the Gaussian models while in the presence of nonlinear coupling terms it provides a measure for the non-Gaussianity of the equilibrium distribution. The local skewness was found to be subject to large statistical uncertainties in the studied finite-size systems with charge conservation and as a consequence will not be discussed in this work.

\begin{figure}
 \centering
 \includegraphics[width=0.48\textwidth]{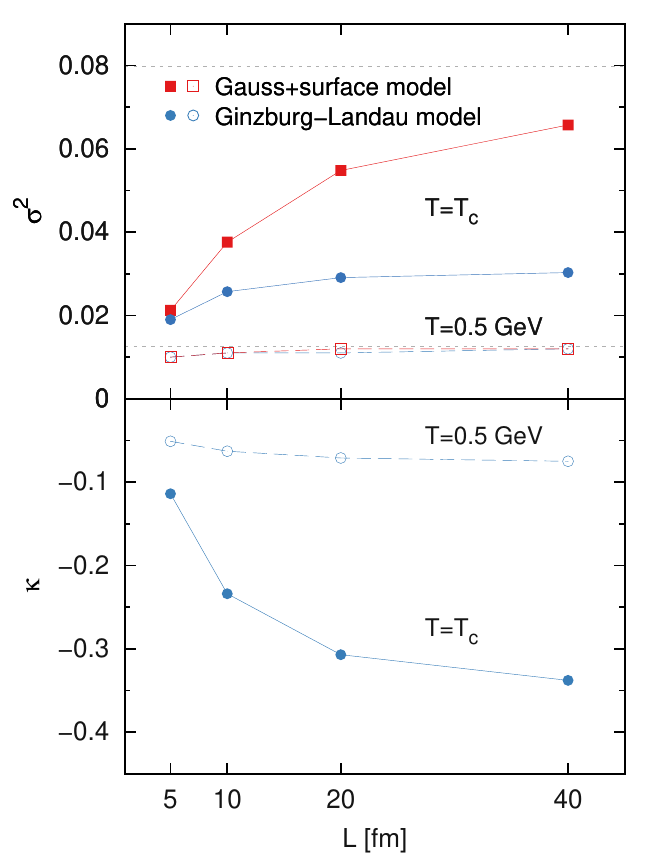}
 \caption{Results for the local variance $\sigma^2$ and local excess kurtosis $\kappa$ in the Gauss$+$surface (squares) and Ginzburg-Landau (circles) models for different system sizes $L$ at fixed resolution $\Delta x=(20/256)$~fm for two different temperatures, $T=0.5$~GeV (open symbols) and $T=T_c$ (solid symbols). The horizontal, grey dotted lines in the upper panel show for comparison the corresponding continuum expectations for $\sigma^2$ in the Gauss$+$surface model for an infinite system, cf.~Eq.~\eqref{equ:LimitLocVariance}.}
 \label{fig:finitesize}
\end{figure}
In Fig.~\ref{fig:finitesize} we show numerical results for the system-size dependence of $\sigma^2$ and $\kappa$ in the Gauss$+$surface and Ginzburg-Landau models for two different $T$ at fixed resolution $\Delta x=(20/256)$~fm. In the Gauss$+$surface model the continuum expectation of $\sigma^2$ for an infinite system is given by 
\begin{equation}
 \label{equ:LimitLocVariance}
 \sigma^2 = \frac{n_c^2}{2A\sqrt{K}m} \,,
\end{equation}
which is indicated by the grey dotted lines. In a finite-size system the local variance can be significantly smaller due to charge conservation, cf.~Fig.~\ref{fig:surfaceCorrFunc3}, but increases with increasing $L$ approaching the limit Eq.~\eqref{equ:LimitLocVariance}. The observed reduction of $\sigma^2$ in the Ginzburg-Landau model is in line with the behavior seen in $m_{\rm eff}$ and $\tilde\xi$, see Fig.~\ref{fig:surfaceCorrFunc} and the discussion in section~\ref{sec:StaticNonGauss}. We find a negligible residual $\Delta x$-dependence in $\sigma^2$ for all $T$ and $L$ similar to $\tilde\xi$. This is in contrast to the behavior noted in~\cite{Nahrgang:2017hkh} for the pure Gaussian model where the local variance depends explicitly on the resolution, cf.~Eq.~\eqref{equ:pureGaussSigma}. This unphysical behavior is cured by the inclusion of a finite surface tension, see also the discussion in~\cite{Bluhm:2019yfb}. The local excess kurtosis vanishes within the statistical uncertainty in the Gauss$+$surface model. In the Ginzburg-Landau model, instead, $\kappa$ is non-zero and found to increase in magnitude with $L$ but also seems to approach a limiting value with increasing system-size. The residual $\Delta x$-dependence is a bit stronger than for $\sigma^2$ but still on the few-percent level. Both $\sigma^2$ and $\kappa$ are significantly larger at $T=T_c$ (solid squares and circles) than at $T=0.5$~GeV (open squares and circles), where the influence of the Gaussian mass parameter is expected to dominate. Near $T_c$ finite-size effects in both observables are clearly more pronounced than at $T=0.5$~GeV and appear to be quantitatively stronger in the higher-order fluctuation observable $\kappa$.

\begin{figure}
 \centering
 \includegraphics[width=0.48\textwidth]{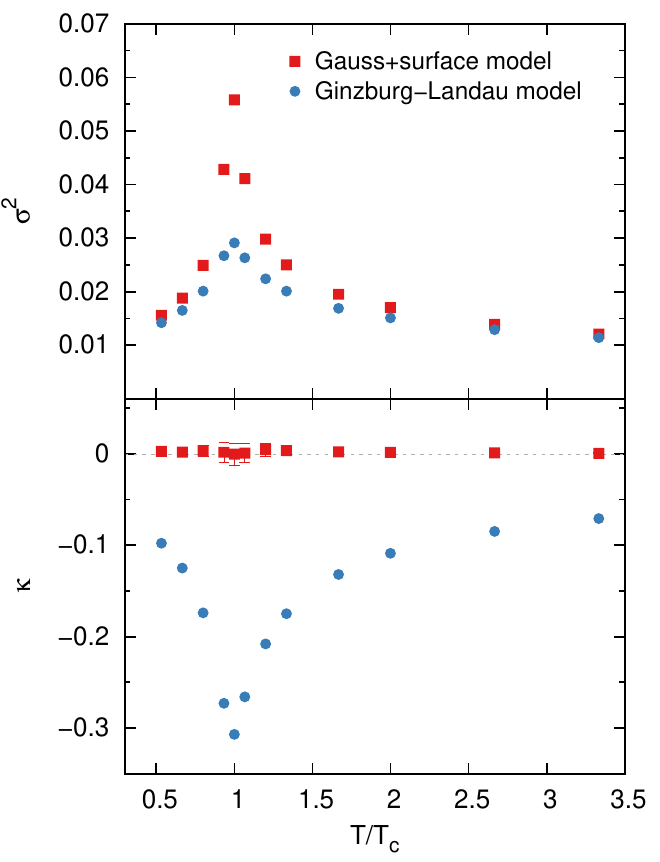}
 \caption{Results for the local variance $\sigma^2$ and local excess kurtosis $\kappa$ as functions of the scaled temperature $T/T_c$ in the Gauss$+$surface (squares) and Ginzburg-Landau (circles) models for simulations with $L=20$~fm and $\Delta x=L/256$.}
 \label{fig:tempdep1}
\end{figure}
In Fig.~\ref{fig:tempdep1} we compare the temperature dependence of the local variance and local excess kurtosis in the Gauss$+$surface (squares) and Ginzburg-Landau (circles) models for $L=20$~fm with $N_x=256$. The reduction seen in $\sigma^2$ for the Ginzburg-Landau model compared to the Gauss$+$surface model is in line with the findings for the temperature dependence of the fitted correlation length in Fig.~\ref{fig:XiFit}. In fact, within the numerics we find that $\sigma^2$ scales approximately as $\sigma^2\sim\tilde{\xi}$ for all $T$ as expected from Eq.~\eqref{equ:LimitLocVariance}. The numerical results for the local excess kurtosis highlight an important difference between the two models: while $\kappa$ vanishes within the acquired statistics in the Gauss$+$surface model, it is non-zero and negative for the chosen values of $\tilde\lambda_i$ in the Ginzburg-Landau model. One observes a non-monotonic temperature dependence with a prominent peak structure in the vicinity of $T_c$, where $\tilde\lambda_4$ and $\tilde\lambda_6$ become the dominant parameters, cf.~Fig.~\ref{fig:parameter}. 

\section{Dynamics of Gaussian and non-Gaussian fluctuations}
\label{sec:dynamics}

We now turn to the study of the dynamics of the system, which we discuss in three steps: first, we investigate the dynamical properties in equilibrium in form of the dynamic structure factor, next we study the response of the system to a sudden quench in temperature and finally look at a Hubble-like reduction of the temperature as a function of time. Note that the dynamical properties depend on the value and/or the temporal behavior of the diffusion coefficient $D$, which as a function of temperature is defined as $D=\Gamma T/n_c$, where we fix $D(T_0)=1$~fm at $T=T_0=0.5$~GeV unless otherwise specified.

\subsection{Dynamic structure factor and relaxation time}
\label{sec:dynamicSF}

The dynamical properties of the system in equilibrium at a fixed temperature are encoded in the dynamic structure factor. The time-dependence of the spatial spectrum of the fluctuating net-baryon density is related to the spatial Fourier transform of the stochastic diffusion equation, Eq.~(\ref{eq:stochdiff}), and can be obtained from $S(k,\omega)$ by the Fourier transformation into the time-domain viz 
\begin{align}
\nonumber
 S(k,t)\equiv & \,\,V\langle \Delta \hat{n}_B(k,t')\,\Delta \hat{n}_B^*(k,t'+t)\rangle \\
 = & \,\,\frac{1}{2\pi} \int_{-\infty}^{\infty} S(k,\omega) \, e^{i\omega t} d\omega \,.
\label{equ:dynamicSdefinitionMixed}
\end{align}
For the Gaussian models with $S(k,\omega)$ given in Eq.~\eqref{equ:dynamicSdefinition} this amounts to 
\begin{equation}
 \label{equ:SkdynMixedCont}
 S(k,t) = S(k) \exp\left(-t/\tau_k\right)
\end{equation}
in the continuum limit, where the static structure factor $S(k)$ is given by Eqs.~\eqref{eq:SkGausscont} or~\eqref{eq:SkGaKcont} and the inverse relaxation time reads 
\begin{equation}
\label{eq:RelaxTimeGAs}
 \tau_k^{-1} = \frac{Dm^2}{n_c}\left(1+\frac{K}{m^2}k^2\right)k^2 \,.
\end{equation}
By setting $K=0$ we find the expression of $\tau_k$ for the pure Gaussian model.

\begin{figure}
 \centering
 \includegraphics[width=0.48\textwidth]{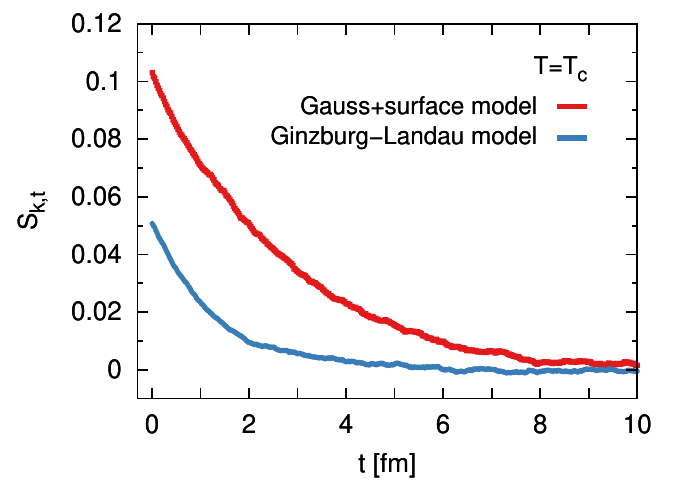}
 \caption{The dynamic structure factor $S_{k,t}$ as a function of time for $\kappa=2$ in the Gauss$+$surface and Ginzburg-Landau models for $L=20$~fm and $N_x=256$ at $T=T_c$.}
 \label{fig:SktTc}
\end{figure}
Numerically, we study the dynamic structure factor $S_{k,t}$ in discretized space-time by analyzing the correlator of the density fluctuations in the mixed representation for modes with given wavevector $\vec{k}$ and wavenumber $\kappa=kL/(2\pi)$, see Appendix~\ref{app:D}. Exemplarily for $\kappa=2$, we contrast $S_{k,t}$ at $T=T_c$ for the Gauss$+$surface 
and Ginzburg-Landau models in Fig.~\ref{fig:SktTc}. One clearly observes an exponential decay of the correlator in both models similar to the expected behavior in the continuum limit. 
As for the static observables, the nonlinear interactions in the Ginzburg-Landau model reduce the dynamic structure factor compared to the Gauss$+$surface model and, in addition, accelerate its exponential decay. We note that in the pure Gaussian model $S_{k,t}$ for the same $\kappa$ is much larger and relaxes significantly slower than in the other models.

\begin{figure}
 \centering
 \includegraphics[width=0.48\textwidth]{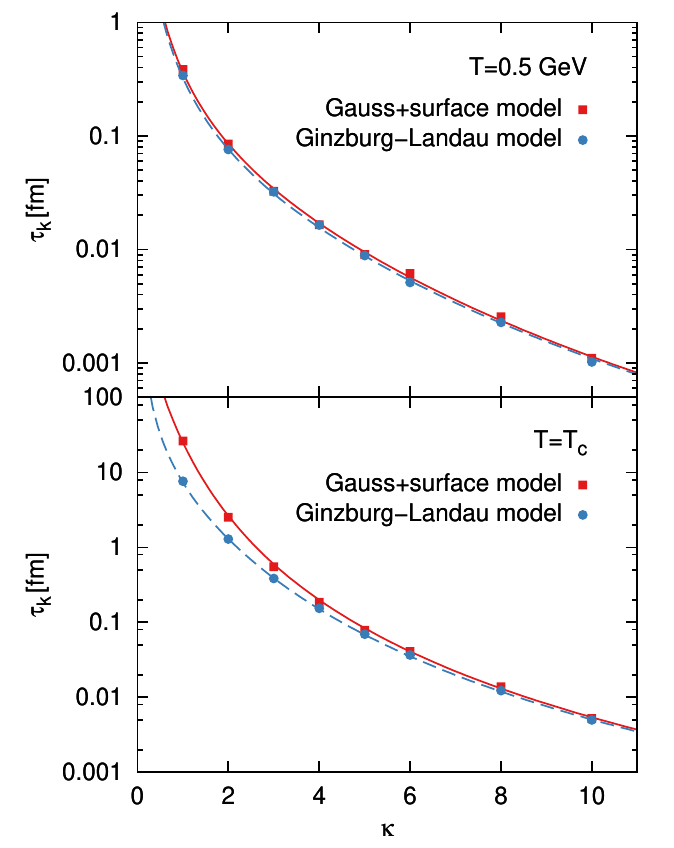}
 \caption{Relaxation time $\tau_k$ (symbols) as a function of $\kappa$ for $L=20$~fm and $N_x=256$ at $T=0.5$~GeV (upper panel) and $T=T_c$ (lower panel) in the Gauss$+$surface and Ginzburg-Landau models. The results are compared with the continuum expression of the Gauss$+$surface model, Eq.~(\ref{eq:RelaxTimeGAs}), shown as red solid lines, and a modified expression with $m$ replaced by $m_{\rm eff}$, see Fig.~\ref{fig:staticSkCorrGL}, shown as blue dashed lines.}
 \label{fig:tauGLvsGAs}
\end{figure}
The relaxation time in the Gauss$+$surface model for a specific mode $k$ can be determined by fitting the corresponding $S_{k,t}$ with an exponential ansatz of the form of the continuum expression. For $T=0.5$~GeV and $T=T_c$ the results for not too large $\kappa$ are shown in Fig.~{\ref{fig:tauGLvsGAs}} (red squares). As one would expect, $\tau_k$ is drastically enhanced near $T_c$ and long-wavelength (small $\kappa$) modes relax significantly slower than short-wave (large $\kappa$) fluctuations. 
The continuum results based on Eq.~(\ref{eq:RelaxTimeGAs}) are also shown as red solid lines in 
Fig.~{\ref{fig:tauGLvsGAs}}. We find that the results of the fits to the data from simulations with $\Delta x = (20/256)$~fm are already very close to the continuum expectations for not too large $\kappa$ (see the discussion in Appendix~\ref{app:D}).

The exponential decay of $S_{k,t}$ seen in Fig.~\ref{fig:SktTc} for the Ginzburg-Landau model suggests to use a similar ansatz to determine $\tau_k$ in this case. The results are shown by blue circles in Fig.~\ref{fig:tauGLvsGAs}. The nonlinear interactions are found to reduce the fitted relaxation time, in particular, for modes with small $\kappa$, and the effect is more prominent in the vicinity of $T_c$. For larger $\kappa$, fluctuations are less affected by the nonlinear interactions and $\tau_k$ in the Gauss$+$surface and the Ginzburg-Landau model is comparable. The $k$-dependence of our numerical results for $\tau_k$ in the Ginzburg-Landau model can quite accurately be described by the continuum expression Eq.~(\ref{eq:RelaxTimeGAs}) of the Gauss$+$surface model by replacing $m$ with $m_{\rm eff}$, see blue dashed lines in Fig.~\ref{fig:tauGLvsGAs}. The values for the modified Gaussian mass parameter $m_{\rm eff}>m$ are those necessary for describing the behavior of the static structure factor in the Ginzburg-Landau model discussed in section~\ref{sec:StaticNonGauss}.

\begin{figure}
 \centering
 \includegraphics[width=0.48\textwidth]{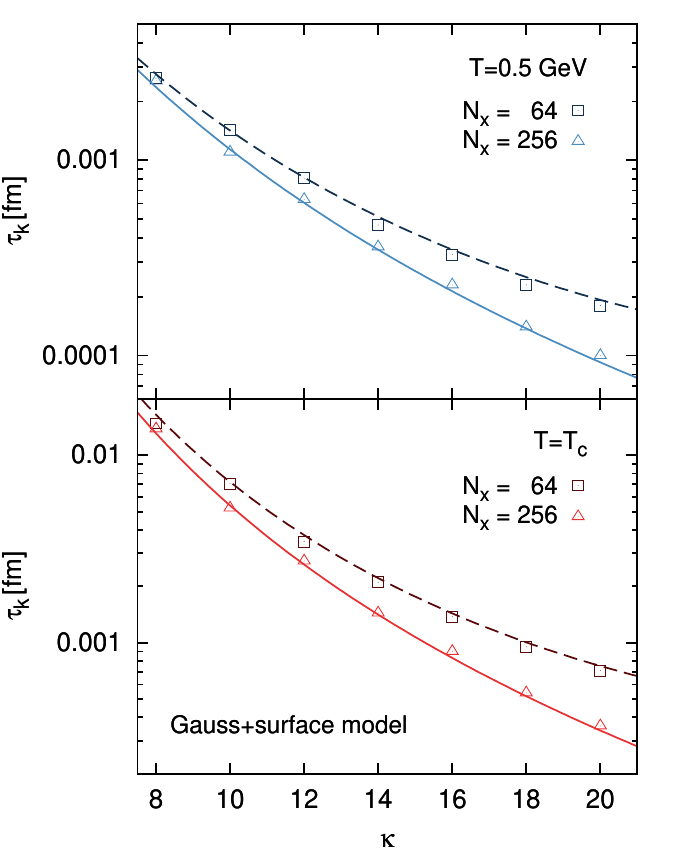}
 \caption{Dependence of the relaxation time (symbols) on the resolution in the Gauss$+$surface model ($\lambda_3=\lambda_4=\lambda_6=0$). We contrast for $T=0.5$~GeV (upper panel) and $T=T_c$ (lower panel) simulations for $N_x=64$ (squares) and $N_x=256$ (triangles) at fixed $L=20$~fm. Analytic expectations based on Eqs.~(\ref{equ:RelaxTimeGAsDisc})~-~\eqref{equ:RelaxTimeGAsDiscEnd} are shown as dashed and solid lines, respectively. We note that the analytic results based on Eq.~(\ref{eq:RelaxTimeGAs}) and on Eqs.~(\ref{equ:RelaxTimeGAsDisc})~-~\eqref{equ:RelaxTimeGAsDiscEnd} are practically indistinguishable for $\Delta x = (20/256)$~fm in the shown range of $\kappa$.}
 \label{fig:tauGAsResolution}
\end{figure}

The comparison of the fit results with the analytic expectations in the Gauss$+$surface model indicates that the simulations carried out with $N_x=256$ at $L=20$~fm are already sufficiently close to the continuum limit, also for the dynamic observables. 
To test further how well analytic expectations for resolution effects are reproduced numerically, we decrease the resolution in the simulations by a factor $4$ and consider $N_x=64$. From Eqs.~(\ref{equ:RelaxTimeGAsDisc})~-~\eqref{equ:RelaxTimeGAsDiscEnd} one expects that a decrease in resolution results in an increase of the fitted relaxation time, in particular for larger $\kappa$. This is precisely observed in the results depicted in Fig.~\ref{fig:tauGAsResolution}. In fact, the comparison of the fit results for $\tau_k$ with the expectations for the relaxation time, Eqs.~(\ref{equ:RelaxTimeGAsDisc})~-~\eqref{equ:RelaxTimeGAsDiscEnd}, shows that resolution effects are well controlled.

\begin{figure}
 \centering
 \includegraphics[width=0.48\textwidth]{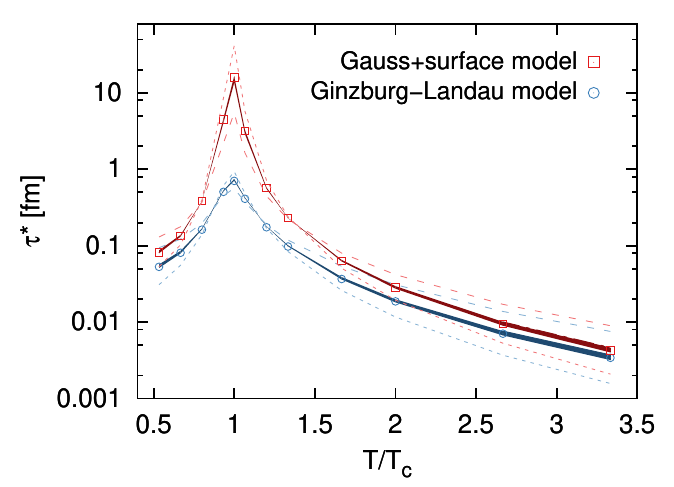}
 \caption{Scaling behavior of the relaxation time $\tau^*$ for modes with $k^*=1/\tilde\xi$ with 
 the correlation length $\tilde\xi$ in Gauss$+$surface and Ginzburg-Landau models as a function of $T/T_c$. Fit results to simulations with $L=20$~fm and $N_x=256$ (symbols) are contrasted with the scaling ansatz $a\tilde\xi^z$, where $a\simeq 0.08/(D\,{\rm fm}^{z-2})$ and $z=4\pm 0.1$ (filled bands) gives the best fit, and $z=3$ (dashed lines) and $z=5$ (dotted lines) are also indicated.}
 \label{fig:RelaxationTimeScaling}
\end{figure}

The determination of the dynamic structure factor and of the relaxation time allows us to study the correlation length dependence of $\tau_k$ for modes which are correlated over the distance $\tilde\xi$. For this purpose, we analyze $\tau^*$, the relaxation time of modes with $k^*=1/\tilde\xi(T)$, as a function of $T$, where $\tilde\xi(T)$ is the fitted correlation length discussed in section~\ref{sec:NumCorrLength}. Results for the Gauss$+$surface and Ginzburg-Landau models are shown in Fig.~\ref{fig:RelaxationTimeScaling} (symbols). We find $\tau^*$ to behave like $a\tilde\xi^z$ with proportionality factor $a$ and dynamic scaling (critical) exponent $z$. For both models, the best fit (filled bands in Fig.~\ref{fig:RelaxationTimeScaling}) gives $z=4\pm 0.1$ and $a\simeq 0.08/(D\,{\rm fm}^{z-2})$. This proportionality factor confirms our expectations, $a=n_c\xi_0/(D(1+\tilde{K}))$, based on the continuum expression of $\tau_k$ in the Gauss$+$surface model. We also indicate that other scaling exponents, e.g.~$z=3$ (dashed lines) or $z=5$ (dotted lines), fail to describe the numerically realized scaling with the correlation length. This shows that our simulations reproduce the dynamic scaling behavior one would expect for the stochastic diffusion of a conserved charge which is the one of model B within the classification scheme~\cite{Hohenberg:1977ym}. 

\subsection{Relaxation of fluctuation observables after a temperature-quench}
\label{sec:QuenchEvolution}

The relaxation dynamics of fluctuation observables such as the local variance $\sigma^2$ and the local excess kurtosis $\kappa$ toward equilibrium can be studied through the sudden quench in temperature from a well prepared initial condition. For this purpose, we first let the system equilibrate at $T=T_0=0.5$~GeV and then instantaneously reduce the temperature at time $\tau=\tau_0$ to three distinct values $T^*$, namely $T^*=T_c$, $T^*=0.18$~GeV and $T^*=0.2$~GeV. We discuss these three quench scenarios only for the Ginzburg-Landau model. Qualitatively, the same conclusions can be drawn for $\sigma^2$ in the Gauss$+$surface model.

\begin{figure}
 \centering
 \includegraphics[width=0.48\textwidth]{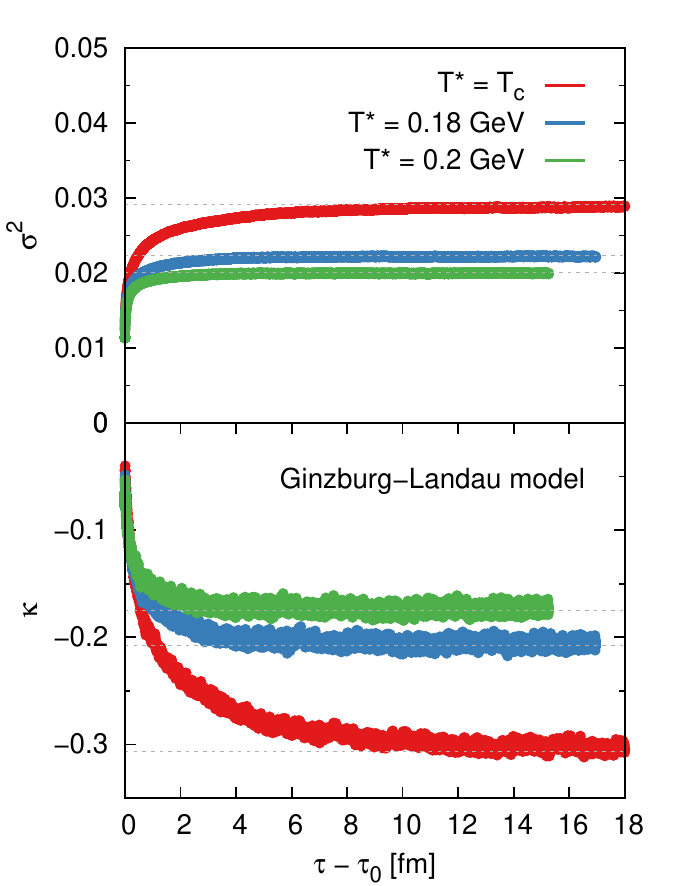}
 \caption{Relaxation dynamics of the local variance $\sigma^2$ and the local excess kurtosis $\kappa$ toward their equilibrium values under a sudden quench in temperature from equilibrium at $T_0=0.5$~GeV to $T^*$ at time $\tau=\tau_0$. The shown results are for the Ginzburg-Landau model with $L=20$~fm and $N_x=256$.}
 \label{fig:quench}
\end{figure}
The results for the relaxation behavior of $\sigma^2$ and $\kappa$ are shown in Fig.~\ref{fig:quench}. One observes that the relaxation dynamics is quite abrupt initially. We find that with decreasing quench-temperature $T^*$ the time it takes $\sigma^2$ and $\kappa$ to relax to the corresponding equilibrium result (horizontal, grey dotted lines) increases. This is to be expected since for smaller $T^*$ we have a smaller diffusion coefficient $D$ and, moreover, the difference between the equilibrium values at $T_0$ and at a $T^*$ close to $T_c$ is larger. In addition, higher moments appear to approach their equilibrium expectations slower. By increasing the initial value of the diffusion coefficient to $D(T=T_0)=2$~fm, the relaxation rate is overall increased and the fluctuation observables relax quicker toward equilibrium, see also the discussions in~\cite{Nahrgang:2017hkh,Bluhm:2019yfb}. We note that the determination of the relaxation time of fluctuation observables within a quench scenario can allow the identification of structures in the QCD phase diagram. This was demonstrated in a QCD-assisted transport approach based on nonequilibrium chiral fluid dynamics and the effective action of low-energy QCD in~\cite{Bluhm:2018qkf}.

\subsection{Time-evolution of fluctuations in a cooling system}
\label{sec:dynamicTimeEvolution}

\begin{figure}
 \centering
 \includegraphics[width=0.48\textwidth]{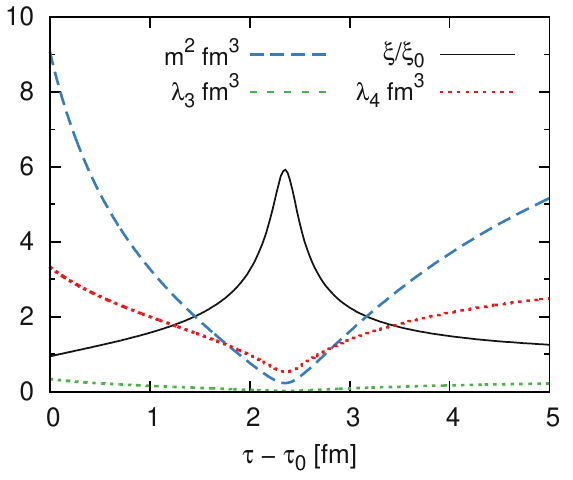}
 \caption{Time-dependence of the parameters entering the Ginzburg-Landau free energy functional ${\cal F}$ in Eq.~(\ref{eq:GLpotential}), see Fig.~\ref{fig:parameter} for comparison. The coupling $\lambda_6=1/{\rm fm}^3$ (not shown) is set constant. The critical temperature $T_c=0.15$~GeV is reached at $\tau_c-\tau_0=2.33$~fm.}
 \label{fig:timeparam}
\end{figure}

Assuming a dynamical evolution of the temperature of the system allows us to highlight some important nonequilibrium effects. For this purpose, we consider a simple, spatially homogeneous time-dependence of $T$ in the Hubble-like form 
\begin{equation}
 T(\tau)=T_0\left(\frac{\tau_0}{\tau}\right)^{dc_s^2}
 \label{equ:Ttimedependence}
\end{equation}
with dimension $d=3$, speed of sound $c_s^2=1/3$ and $T_0=0.5$~GeV at initial time $\tau_0=1$~fm at which the system is in equilibrium. For this cooling scenario the critical temperature is reached at $\tau_c-\tau_0=2.33$~fm. The time-dependent temperature translates into time-dependent couplings via Eqs.~(\ref{eq:couplingsA})~-~(\ref{eq:couplings}), which are shown in Fig.~\ref{fig:timeparam}. Due to the fast initial decrease of $T$ and the slower decrease at later times in Eq.~\eqref{equ:Ttimedependence} the thermodynamic correlation length is more symmetric between the early and late times than it is in comparison to the high and low temperatures in Fig.~\ref{fig:parameter}. Still, all couplings except $\lambda_6$, which is independent of the correlation length, have a dip at the time when the critical temperature is reached. This is the region where we expect nonequilibrium effects to be most prominent.

\begin{figure}
 \centering
 \includegraphics[width=0.48\textwidth]{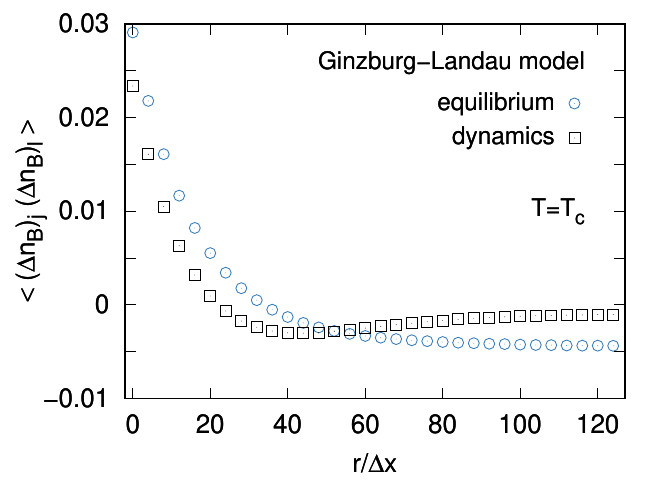}
 \includegraphics[width=0.49\textwidth]{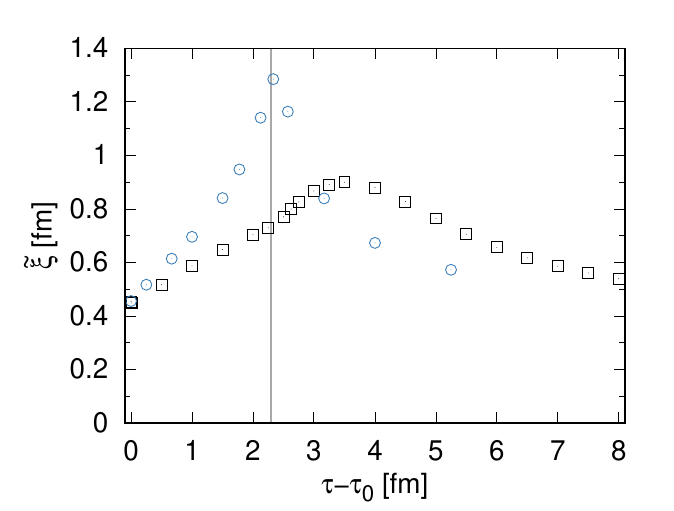}
 \caption{Dynamical behavior (squares) of the equal-time correlation function (upper panel) and the fitted correlation length (lower panel) in a cooling system, cf.~Eq.~\eqref{equ:Ttimedependence}, in comparison with the equilibrium situation (circles) at the same $T$. The results are obtained for the Ginzburg-Landau model with $L=20$~fm and $N_x=256$. We show a snapshot of the correlation function as a function of distance $r=|j-l|\Delta x$ with $\Delta x=L/N_x$ at time $\tau_c-\tau_0=2.33$~fm, i.e.~at $T=T_c$, and $\tilde\xi$ as a function of $\tau-\tau_0$, where the vertical line indicates $\tau_c-\tau_0$.}
 \label{fig:dynamicCorrFunc}
\end{figure}

We first study the impact of the dynamical evolution of $T$ on the equal-time correlation function and the associated correlation length $\tilde\xi$. The results presented here are obtained for the Ginzburg-Landau model where we analyzed a sufficient amount of events. The form of the equal-time correlation function is clearly affected by the dynamics, see upper panel in Fig.~\ref{fig:dynamicCorrFunc} (squares) for $T=T_c$. On the quantitative level, this is also determined by the temporal evolution of the diffusion coefficient $D$. For not too large initial values (such as $D(T_0)=1$~fm) it already significantly decreased (to $D(T_c)=0.3$~fm in this case) by the time $T_c$ is reached and, thus, local fluctuations cannot rapidly enough be balanced throughout the entire finite-size system. As a consequence, correlations at zero distance do not build up quickly enough from the smaller value at $T_0$ toward the equilibrium value at $T_c$ (see upper panel in Fig.~\ref{fig:dynamicCorrFunc} (circles) and Fig.~\ref{fig:tempdep1}) and lag behind. Around these local fluctuations, anticorrelations are present due to charge conservation. In the dynamical situation they do not have sufficient time to diffuse into the entire system. We therefore see a dip of the correlation function around $r=40\,\Delta x$, while it approaches zero at larger distances. This local balancing of the fluctuations reduces the correlation length, as we discuss in the following.

While the form of the equal-time correlation function in the dynamical scenario is not in one-to-one correspondence with the equilibrium form, we may still analyze the visible exponential decrease in the region from small to intermediate distances to deduce a correlation length. By using the ansatz employed in section~\ref{sec:NumCorrLength} for different $T$, i.e.~at different times $\tau-\tau_0$, we obtain the result for the dynamical $\tilde\xi$ shown in the lower panel of Fig.~\ref{fig:dynamicCorrFunc} (squares). In comparison to the equilibrium result (circles, cf.~also Fig.~\ref{fig:XiFit}) we observe clear deviations highlighting two distinct nonequilibrium effects: first, the overall magnitude of $\tilde\xi$ is significantly reduced as a consequence of the dynamics. Secondly, there are clear indications for a retardation effect due to the rapid cooling in $T$. The dynamical $\tilde\xi$ remains initially smaller than its equilibrium counterpart for given $T$ but then develops a maximum at a temperature far below $T_c$ such that at late times it is actually larger than in the equilibrium situation. The pronounced structure traditionally associated with the phase transition is shifted to later times and, thus, different thermal conditions. We expect similar effects for the fluctuation observables.

\begin{figure}
 \centering
 \includegraphics[width=0.48\textwidth]{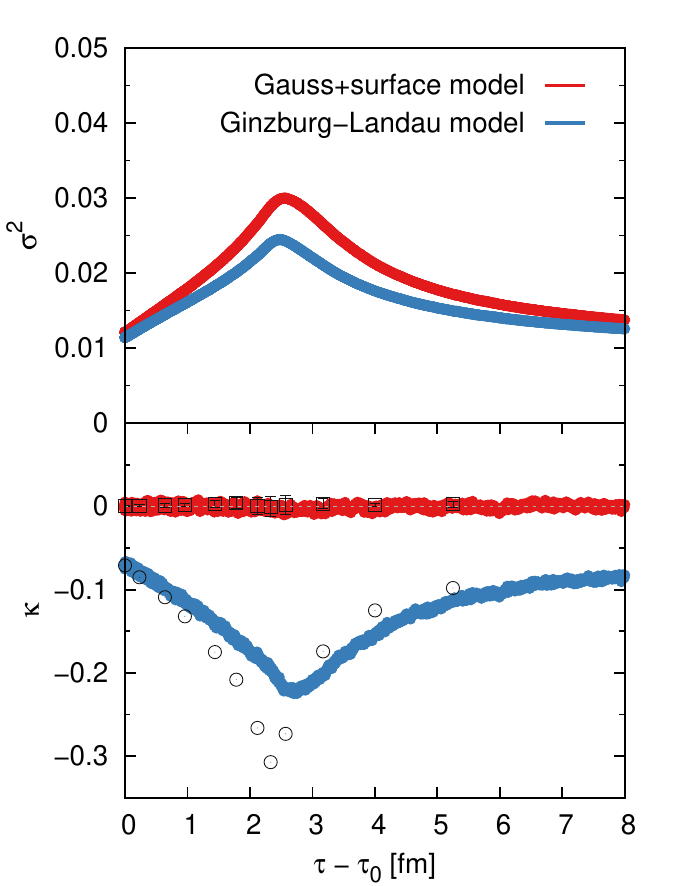}
 \caption{Dynamical behavior (colored bands) of the local variance $\sigma^2$ and the local excess kurtosis $\kappa$ for the Gauss$+$surface and Ginzburg-Landau models in a cooling system, cf.~Eq.~\eqref{equ:Ttimedependence}, as a function of time. For the local excess kurtosis we compare with the equilibrium results (open squares and circles) at the corresponding $T$. The shown results are for $L=20$~fm and $N_x=256$.}
 \label{fig:dynamic}
\end{figure}

In Fig.~\ref{fig:dynamic} we show the temporal evolution of the local variance $\sigma^2$ (upper panel) and the local excess kurtosis $\kappa$ (lower panel) as a function of time in the Gauss$+$surface (red bands) and Ginzburg-Landau (blue bands) models. For both models, $\sigma^2$ peaks at a time $\tau-\tau_0$ shortly after $T_c$ is reached during the evolution. The retardation shift appears slightly larger in the Gauss$+$surface than in the Ginzburg-Landau model. As in the equilibrium situation, $\sigma^2$ in the Ginzburg-Landau model stays below the Gauss+surface model result, but the reduction of its maximal value due to the dynamics is significantly stronger in the Gauss$+$surface model (by 46\% compared to 17\% for the Ginzburg-Landau model). 

For the local excess kurtosis we note that in the absence of nonlinear coupling terms $\kappa$ vanishes in the dynamical scenario as it did in equilibrium (see red band and open squares in the lower panel of Fig.~\ref{fig:dynamic}). For the Ginzburg-Landau model $\kappa$ starts at its equilibrium value for $T_0$ and initially follows the equilibrium behavior for given $T$ (see blue band and open circles in the lower panel of Fig.~\ref{fig:dynamic}). However, within the band of statistical uncertainties it quickly lags behind the equilibrium situation as reflected in the reduced magnitude of $\kappa$. We can clearly see that in the dynamical scenario the minimum in the local excess kurtosis is shifted to a later time than $\tau_c-\tau_0$ and that the magnitude of this minimum is significantly reduced (by approximately $30$\%) compared to the equilibrium result. At later times, the retardation effect leads to a dynamical $\kappa$ slightly larger in magnitude than in equilibrium. As is evident from Fig.~\ref{fig:dynamic}, the nonequilibrium effects influence $\kappa$ stronger than $\sigma^2$ in the Ginzburg-Landau model. The behavior seen in the fluctuation observables resembles qualitatively the one discussed for $\tilde\xi$ above with an important difference: the shift of the maximum in $\tilde\xi$ to smaller $T$ is larger than in $\sigma^2$ or $\kappa$. In future work we will investigate the fluctuation observables over larger subvolumes to see if the relation with the correlation length is restored. We note that an overall reduction of the dynamical diffusion coefficient (by lowering its initial value $D(T_0)$) results in a stronger retardation and a stronger reduction of the magnitude of the fluctuation signal.

\section{Discussion}
\label{sec:summary}

In this work we presented a first rigorous implementation of the one-dimensional stochastic diffusion equation near the QCD critical point. First, we benchmarked this implementation in the linear approximation, including Gaussian mass and surface tension terms, versus analytic results of the equal-time correlation function and the static structure factor. Based on these tests, we chose the resolution of the spatial discretization as to reproduce the behavior of the first $50$\% of the wavenumbers in the continuum limit. Charge conservation is found to play an important role for the correlation function and limits the growth of long-range correlations. In the same sense the growth of the correlation length near $T_c$ is limited for system sizes up to a few times the thermodynamic correlation length.

In equilibrium we investigated the temperature dependence of the local variance and excess kurtosis. The latter takes non-zero values as soon as the nonlinear coupling terms in the Ginzburg-Landau free energy  functional are included. The expected non-monotonic behavior around $T_c$ is clearly observed. The inclusion of the nonlinear coupling terms reduces the variance of the system by a factor of two near $T_c$.

From the dynamic structure factor we obtained the relaxation time of the critical mode. It is found to scale with the correlation length according to model B of dynamical universality. Finally we investigated the response of the system to changes of the temperature, first, via a sudden quench in temperature and, second, via a Hubble-like time evolution. We observe again that the growth of the correlation length is limited by charge conservation effects, this time in a dynamical setup. Here, fluctuations do not have enough time to diffuse to larger distances and, thus, the correlation length is limited to a smaller range. Fluctuation observables are reduced in magnitude and shifted to smaller temperatures due to nonequilibrium effects. Higher-order cumulants are impacted stronger than the variance by the nonequilibrium situation, i.e.~they need more time to relax, the magnitude of their extrema is more reduced compared to the equilibrium values and the retardation effect is stronger. 

We emphasize in particular the importance of benchmarking the approach to the dynamics of fluctuations against analytic results, like the correlation function, the static and the dynamic structure factor. This should be a standard requirement for all models dealing with the dynamics of fluctuations, including more complex approaches to fluctuating fluid dynamics.

The presented cumulants are evaluated as local observables over individual cells of the simulation region, which serves well the purpose of understanding the basic dynamics of fluctuations in stochastic partial differential equations. For aiming at a comparison with experimental data from heavy-ion collisions, integrated observables in finite kinematic regions are of additional interest. A study of fluctuations over larger subregions of observation similar to~\cite{Nahrgang:2018afz} and of their systematics will address these questions and be reported elsewhere. In our studies of the time evolution of the temperature the considered systems did not expand. We plan to investigate the expansion of the system in a next step, see~\cite{Kitazawa:2020kvc}, to include regular contributions into the free energy functional and to extend the treatment of fluctuations to three spatial dimensions.

\section*{Acknowledgements}
The authors acknowledge the support of the program ``Etoiles montantes en Pays de la Loire 2017''. This research was supported in part by the ExtreMe Matter Institute (EMMI) at the GSI Helmholtzzentrum f\"ur Schwerionenforschung, Darmstadt, Germany. The authors thank S.~A.~Bass and T.~Sch\"afer for stimulating discussions.

\appendix

\section{Numerical implementation}
\label{app:A}

We study the stochastic diffusion equation Eq.~\eqref{eq:stochdiff} by discretizing the diffusive net-baryon density on $N_x$ sites equally distributed over a system of longitudinal extent $L$ with resolution (grid spacing) $\Delta x=L/N_x$. Over the finite $\Delta x$ of a cell, $n_B$ must be understood as being averaged. Time is discretized in steps of $\Delta t$ at which $n_B$ is considered point-wise. Details about the extent and discretization in the transverse direction are not important as we study the evolution of the system and physical observables only in the longitudinal direction which is decoupled from the transverse dynamics. For simplicity we set the transverse area $A=1$~fm$^2$ but have verified the proper behavior with $A$ in the numerics. The Gaussian white noise must be understood as averaged over space $\Delta x$ and time $\Delta t$. It is independent between different cells and time steps with zero mean and variance $1/(\Delta x\Delta t)$.

Equation~\eqref{eq:stochdiff} is solved by means of a semi-implicit scheme. While the operator associated with the Gaussian mass and surface tension terms is treated implicitly in time, the operator associated with the nonlinear coupling terms is discretized explicitly. The temporal integration is performed with a predictor-corrector method. For the stochastic diffusion equation of the general form 
\begin{equation}
 \frac{dn_B}{dt} = {\cal O}_l n_B + {\cal O}_{nl}(n_B) + {\cal O}_\xi W
\end{equation}
this amounts to solving in a first step 
\begin{align}
\nonumber
 \left(1-\frac{\Delta t}{2}{\cal O}_l\right) \tilde n_B^{m+1} = & \left(1+\frac{\Delta t}{2}{\cal O}_l\right) n_B^m + \Delta t \,{\cal O}_{nl}(n_B^m) \\
 & + \Delta t \,{\cal O}_\xi W^m
\end{align}
for $\tilde n_B^{m+1}$ as an intermediate update of $n_B^m$ from timestep $m$, and then by using $n_B^m$ and $\tilde n_B^{m+1}$ in 
\begin{align}
\nonumber
 \left(1-\frac{\Delta t}{2}{\cal O}_l\right) n_B^{m+1} = & \left(1+\frac{\Delta t}{2}{\cal O}_l\right) n_B^m \\
\nonumber
 & + \frac{\Delta t}{2} \left[{\cal O}_{nl}(\tilde n_B^m)+{\cal O}_{nl}(n_B^m)\right] \\
 & + \Delta t \,{\cal O}_\xi W^m
\end{align}
one finds $n_B^{m+1}$ as the update at timestep $m+1$. The individual operators in the above equations read (we drop the index $B$ in the following to improve readability)
\begin{align}
\nonumber
 {\cal O}_l n^m & = \frac{D}{\Delta x^2} \frac{m^2}{n_c} \left(n^m_{j+1}-2n^m_{j}+n^m_{j-1}\right) \\
 - \frac{D}{\Delta x^4} \frac{K}{n_c} &  \left(n^m_{j+2}-4n^m_{j+1}+6n^m_{j}-4n^m_{j-1}+n^m_{j-2}\right) , \\
\nonumber
 {\cal O}_{nl}(n^m) & = \frac{D}{\Delta x^2} \sum_{i=3,4,6} \frac{\lambda_i}{n_c^{i-1}} \Big[\left(\Delta n_{j+1}^m\right)^{i-1} \\
\label{eq:NonlinearOp}
 & -2\left(\Delta n_{j}^m\right)^{i-1}+\left(\Delta n_{j-1}^m\right)^{i-1}\Big] , \\
 {\cal O}_\xi W^m & = \sqrt{\frac{2Dn_c}{A\Delta x\Delta t}} \frac{1}{\Delta x} \left(W^m_{j+\frac12}-W^m_{j-\frac12}\right) .
\label{eq:Noisedisc}
\end{align}
This system of equations is solved for each spatial point $j$ on the grid. The contributions from the nonlinear coupling terms are simulated by computing the corresponding power of $\Delta n^m_j=n^m_j-n_c$ at a given site $j$ for each timestep $m$. Without nonlinear coupling terms the predictor and corrector steps are identical and yield exactly the same solution which makes one of the steps redundant. The noise field $W$ in Eq.~\eqref{eq:Noisedisc} has zero mean and variance $1$.

\section{Static structure factor in discretized space-time}
\label{app:A+}

For the Gauss and Gauss$+$surface models, for which the contributions from the nonlinear operator in Eq.~\eqref{eq:NonlinearOp} vanish, analytic results for the static structure factor $S_k$ in discretized space-time can be derived. In this limit, the general form of the stochastic diffusion equation may be written in mixed Fourier space as
\begin{equation}
\label{eq:AppAmixedFourier}
 M_k^{(1)} \hat{n}_k^{m+1} = M_k^{(-1)} \hat{n}_k^{m} + N_k\,\hat{W}_k^m
\end{equation}
with 
\begin{align}
\nonumber
 M_k^{(a)} = & \,1 - a\frac{\Delta t}{2} \Big(\frac{D}{\Delta x^2}\frac{m^2}{n_c}\left[2\cos(\Delta k)-2\right] \\
\label{eq:MkinMixedFourier}
 & \,-\frac{D}{\Delta x^4}\frac{K}{n_c}\left[2\cos(2\Delta k)-8\cos(\Delta k)+6\right]
 \Big) \,, \\
\label{eq:NkinMixedFourier}
 N_k = & \,\sqrt{\frac{8Dn_c\Delta t}{A\Delta x^3}}\sin(\Delta k/2) \,,
\end{align}
$\Delta k=k\Delta x$ and $\langle\hat{W}_k^m (\hat{W}_k^m)^*\rangle=1/N_x$. We note that based on Eq.~\eqref{eq:stochdiff} we find that Eq.~\eqref{eq:AppAmixedFourier} holds true also for the difference $\Delta\hat{n}_k$ instead of $\hat{n}_k$. From the definition 
\begin{equation}
\label{eq:StaticStrucFacDef}
 S_k^m = V \langle\Delta\hat{n}_k^m (\Delta\hat{n}_k^m)^*\rangle
\end{equation}
and the condition of stationarity $S_k = S_k^m = S_k^{m+1}$ in equilibrium one finds 
\begin{align}
 S_k = & \,\frac{|N_k|^2 A\Delta x}{\left|M_k^{(1)}\right|^2-\left|M_k^{(-1)}\right|^2} \\
\label{eq:SkGAsdiscrete}
 = & \,\frac{n_c^2}{m^2}\frac{1}{1+\frac{2K}{m^2\Delta x^2}\left[1-\cos(\Delta k)\right]}
\end{align}
for the static structure factor, see Eq.~\eqref{equ:StrucFacDiscretGA+K}. This result is independent of the time-step $\Delta t$. In the limit of the pure Gaussian model with $K=0$ this reduces to $S_k=n_c^2/m^2$, which is independent of $k$ and $\Delta x$ and agrees with the result in the continuum, see Eq.~\eqref{eq:SkGausscont}. We note that for $k=0$ the static structure factor reflects charge conservation in the entire system.

\section{Net-baryon number conservation in finite-size systems}
\label{app:B}

The net-baryon number of the entire system is numerically conserved by imposing periodic boundary conditions on the net-baryon number density. As a result, charge conservation must be reflected in the behavior of observables such as the equal-time correlation function. The latter is connected with the static structure factor by a Fourier transformation. In discretized space one defines
\begin{equation}
 n_j = \sum_k \hat{n}_k e^{i j k \Delta x}\,,
\end{equation}
where $k=2\pi\kappa/L$ is restricted by $0\leq \kappa < N_x$ for $\kappa\in {\cal Z}$. Accordingly, the equal-time correlation function follows in discretized space as 
\begin{equation}
 \langle(\Delta n_B)_j(\Delta n_B)_l\rangle = \frac{1}{V} \sum_{\kappa=0}^{N_x-1} e^{i \,2\pi\kappa|j-l|/N_x} 
 S_k \,.
\end{equation}
This definition holds for an infinite system. For the pure Gaussian model with $S_k=n_c^2/m^2$ one finds 
\begin{equation}
\label{eq:CorrGauss}
 \langle(\Delta n_B)_j(\Delta n_B)_l\rangle = \frac{n_c^2}{Am^2}\frac{\delta_{jl}}{\Delta x}
\end{equation}
because modes with different $\kappa$ are orthogonal.

For a finite-size system, however, charge conservation must be imposed by demanding that local fluctuations vanish upon summation over the entire system, i.e. $\sum_l \langle(\Delta n_B)_j(\Delta n_B)_l\rangle = 0$ for any $j$. This is respected if we impose 
\begin{align}
 \nonumber
 \langle(\Delta n_B)_j(\Delta n_B)_l\rangle & = \frac{1}{V} \sum_{\kappa=0}^{N_x-1} e^{i \,2\pi\kappa|j-l|/N_x} 
 S_k \\
\label{eq:CorrfunCorr}
 & - \frac{1}{N_xV} \sum_{h=0}^{N_x-1} \sum_{\kappa=0}^{N_x-1} e^{i \,2\pi\kappa h/N_x} 
 S_k \,.
\end{align}
Instead of Eq.~\eqref{eq:CorrGauss} one finds 
\begin{equation}
 \langle(\Delta n_B)_j(\Delta n_B)_l\rangle = \frac{n_c^2}{Am^2}\left(\frac{\delta_{jl}}{\Delta x}-\frac{1}{L}\right) 
\end{equation}
for the pure Gaussian model. The finite-size correction vanishes in the thermodynamic limit for any given resolution $\Delta x$.

For the Gauss$+$surface model with $S_k$ given in Eq.~\eqref{eq:SkGAsdiscrete} the result of the summations in Eq.~\eqref{eq:CorrfunCorr} can be obtained numerically. Due to Eq.~\eqref{eq:CorrfunCorr} one expects a negative shift in a finite-size system. This shift has to become less pronounced with increasing $N_x$, i.e.~for fixed resolution $\Delta x$ with increasing $L$. Both these features are seen in the numerics, cf.~Fig.~\ref{fig:surfaceCorrFunc3}.

\section{Determination of the correlation length}
\label{app:C}

The form of the equal-time correlation function found in the equilibrium simulations is that of an exponential decay which is modified by a negative shift due to exact charge conservation. Since in equilibrium the system is homogeneous on length scales larger than the noise correlation this shift is expected to be a constant. Then, a suitable ansatz to determine the correlation length $\tilde\xi$ is 
\begin{equation}
\label{equ:ansatz}
 \langle(\Delta n_B)_j(\Delta n_B)_l\rangle = \frac{C_1}{C_2} \exp(-|j-l|\Delta x/C_2) + C_3 \,,
\end{equation}
where $C_2$ is the fit parameter for $\tilde\xi$ in dependence of $\Delta x$, $L$ and $T$. The quantity $C_1/C_2+C_3$ gives the value of the correlation function over distances of the grid spacing $\Delta x$, i.e.~the value of the local variance. The fit results for $\tilde\xi$ shown in Sec.~\ref{sec:NumCorrLength} are obtained by optimizing the description of the local variance in the numerics. We observe that for the Gauss$+$surface model simulations with $L=20$~fm the obtained value of $C_1/C_2$ is already quite close to the continuum expectation of $n_c^2/(2Am\sqrt{K})$ for the local variance in an infinite system, cf.~Eqs.~\eqref{eq:CorrFunction1D} and~\eqref{equ:LimitLocVariance}, even for $T$ near $T_c$. We note that for smaller $L$ this is not necessarily the case, in particular close to $T_c$. Motivated by the fact that the equilibrium results of the equal-time correlation function in the Ginzburg-Landau model can be described by the theoretical expectation of the Gauss$+$surface model with a modified, effective Gaussian mass parameter, see Sec.~\ref{sec:StaticNonGauss}, we utilize the same ansatz Eq.~\eqref{equ:ansatz} and strategy in order to fit the numerical results of the Ginzburg-Landau model and to determine $\tilde\xi$.

\section{Dynamic structure factor in discretized space-time}
\label{app:D}

The diffusion equation in discretized space-time discussed in Appendix~\ref{app:A} has for the Gauss and Gauss$+$surface models the following representation in full $(\omega,k)$ Fourier-space: 
\begin{equation}
\label{eq:DiffDiscFourier}
 M_k^{(1)} e^{i\Delta\omega} \hat{n}_{k,\omega} = M_k^{(-1)} \hat{n}_{k,\omega} + N_k \hat{W}_{k,\omega}
\end{equation}
with $\Delta\omega=\omega\Delta t$. This implies for the correlator 
\begin{align}
\nonumber
 & \langle\Delta\hat{n}_{k,\omega}\Delta\hat{n}_{k,\omega}^*\rangle = \\
 & \hspace{2mm} \frac{N_k\langle\hat{W}_{k,\omega}\hat{W}_{k,\omega}^*\rangle N_k^*}{\left(M_k^{(1)}-e^{-i\Delta\omega}M_k^{(-1)}\right)\left(M_k^{(1)^{\,*}}-e^{i\Delta\omega}M_k^{(-1)^{\,*}}\right)}
\end{align}
which gives the dynamic structure factor via 
\begin{equation}
 S_{k,\omega} = \lim_{N_t\to\infty} V(N_t\Delta t) \langle\Delta\hat{n}_{k,\omega}\Delta\hat{n}_{k,\omega}^*\rangle \,,
\end{equation}
where $N_t$ is the number of (performed) time steps. From this definition it is clear that the dynamic structure factor is a late-time equilibrium observable. For white noise we have $\langle\hat{W}_{k,\omega}\hat{W}_{k,\omega}^*\rangle=1/(N_xN_t)$, and with $M_k^{(a)}$ and $N_k$ defined in Appendix~\ref{app:A+} we obtain 
\begin{equation}
\label{eq:DynStrucFacDisc}
 S_{k,\omega} = 
 \frac{2n_c^3\tilde\chi_1Dk^2}{n_c^2\,\Delta t^{-2}(1-\cos(\Delta\omega))+\tilde\chi_2\tilde\chi_1^2D^2k^4}
\end{equation}
with 
\begin{align}
 \tilde\chi_1 & = (1-\cos(\Delta k))/\Delta k^2 \,, \\
 \tilde\chi_2 & = m^4\left(1-\frac{2K}{m^2\Delta x^2}(\cos(\Delta k)-1)\right)^2 (1+\cos(\Delta\omega)) \,.
 \label{equ:chitilde2}
\end{align}
The result for the pure Gaussian model is found by setting $K=0$ in Eq.~\eqref{equ:chitilde2}. In the limit of small $\Delta\omega\ll 1$ we can expand $\cos(\Delta\omega)$ and find 
\begin{align}
\nonumber
 & \lim_{\Delta\omega\ll 1} S_{k,\omega} = \\
 & \hspace{2mm} \frac{4n_c\tilde\chi_1Dk^2}{\omega^2+4D^2\frac{m^4}{n_c^2}k^4\tilde\chi_1^2 \left(1-\frac{2K}{m^2\Delta x^2}(\cos(\Delta k)-1)\right)^2} \,.
\label{equ:dynSsmallomega}
\end{align}
Moreover, in the limit of small $\Delta k\ll 1$ we have $\tilde\chi_1\approx \frac12 - \frac{1}{24}\Delta k^2$ and $(2\cos(\Delta k)-2)/\Delta x^2\approx -k^2 + \frac{1}{12}k^2\Delta k^2$ in Eq.~\eqref{equ:dynSsmallomega}. Thus, for given $\omega$ and $k$, in the limit of $\Delta t\to 0$ and $\Delta x\to 0$ the continuum expression Eq.~\eqref{equ:dynamicSdefinition} of the dynamic structure factor $S(k,\omega)$ is recovered from $S_{k,\omega}$. In the numerics, finite resolution in $\Delta t$ and $\Delta x$ implies deviations from the continuum result. Therefore, only the regime of small wavevectors and low frequencies allows us to judge the accuracy of the numerical scheme. Even in the limit $\Delta t\to 0$, the approach to the continuum is limited to small values of $k$ depending on the spatial resolution. This limit can be used to determine an analytic expression for the dynamic structure factor $S_{k,t}$ in the mixed representation from the Fourier transformation into the time-domain. We find 
\begin{equation}
 S_{k,t} = \frac{1}{2\pi} \int_{-\infty}^{\infty} e^{i\omega t} \lim_{\Delta t\to 0} S_{k,\omega} \,d\omega = S_k \,e^{-t/\tau_k} \,,
\end{equation}
where $S_k$ is the static structure factor in Eq.~\eqref{eq:SkGAsdiscrete} and $\tau_k$ is the relaxation time of fluctuations with wavevector $k=2\pi\kappa/L$ given via 
\begin{equation}
\label{equ:RelaxTimeGAsDisc}
 \tau_k^{-1} = 2\frac{D}{n_c}m^2 k^2\left(A_k+B_k\right)
\end{equation}
with
\begin{align}
 A_k & = (1-\cos(\Delta k))/\Delta k^2 \,, \\
 B_k & = \frac{2Kk^2(1-\cos(\Delta k))^2}{m^2\Delta k^4} \,.
\label{equ:RelaxTimeGAsDiscEnd}
\end{align}
From Eqs.~\eqref{equ:RelaxTimeGAsDisc}~-~\eqref{equ:RelaxTimeGAsDiscEnd} in the limit of small $\Delta x$ we see that finite-resolution effects increase $\tau_k$ compared to the continuum result Eq.~\eqref{eq:RelaxTimeGAs}, which is approached in the limit $\Delta x\to 0$. Moreover, we find that $\tau_k$ is smaller in the Gauss$+$surface model compared to the pure Gaussian model with $K=0$. This effect is less pronounced for small values of $\kappa$ and away from the transition temperature $T_c$.

\end{document}